\documentclass[graybox]{svmult}

\usepackage{mathptmx}       %
\usepackage{helvet}         %
\usepackage{courier}        %
\usepackage{type1cm}        %
\usepackage{makeidx}         %
\usepackage{graphicx}        %
\usepackage{multicol}        %
\usepackage[bottom]{footmisc}%
\usepackage{soul}            %

\definecolor{linkcolor}{rgb}{0.0,0.3,0.5}
\usepackage[unicode,colorlinks=true, linkcolor=linkcolor, citecolor=linkcolor, filecolor=linkcolor,urlcolor=linkcolor]{hyperref}

\usepackage[square,numbers]{natbib}
\makeindex             %

\usepackage{amsmath}

\usepackage{cleveref}
\usepackage{xspace}
\usepackage{empheq}
\usepackage{aas_macros}

\newcommand{\glname}{Glossary and main symbols}

\newcommand{\msun}{\ensuremath{\mathrm{M}_\odot}}

\newcommand{\beq}{\begin{equation}}
\newcommand{\eeq}{\end{equation}}
\newcommand{\beqa}{\begin{eqnarray}}
\newcommand{\eeqa}{\end{eqnarray}}
\newcommand{\boxit}[2]{\begin{empheq}[box={\fboxsep=6pt\fbox}]{equation}  #1 \label{Eq.#2}\end{empheq}}

\newcommand{\vtheta}{\ensuremath{\vec{\theta}}\xspace}

\newcommand{\vl}{\ensuremath{\vec{\lambda}}\xspace}
\newcommand{\La}{\ensuremath{{\Lambda}}\xspace}

\newcommand{\vt}{\ensuremath{\vec{\theta}}\xspace}
\newcommand{\veps}{\ensuremath{\vec{\epsilon}}\xspace}

\newcommand{\dvt}{\ensuremath{\delta\vec{\theta}}\xspace}
\newcommand{\intvt}{\ensuremath{{\left[\vt,\vt+\dvt\right]}}\xspace}

\newcommand{\dlt}{\ensuremath{\delta t}\xspace}
\newcommand{\ndj}{\ensuremath{{\bar{d}_j}}\xspace}
\newcommand{\di}{\ensuremath{{d_i}}\xspace}

\newcommand{\maH}{\ensuremath{\mathcal{H}}\xspace}
\newcommand{\maHL}{\ensuremath{\mathcal{H}_\La}\xspace}
\newcommand{\maHl}{\ensuremath{\mathcal{H}_{\vl}}\xspace}
\newcommand{\maHP}{\ensuremath{\mathcal{H}_{\mathrm{PE}}}\xspace}

\newcommand{\Nexps}{\ensuremath{{N^{s}}_\mathrm{exp}}\xspace}
\newcommand{\Nexpb}{\ensuremath{{N^{b}}_\mathrm{exp}}\xspace}
\newcommand{\Nacts}{\ensuremath{{N^{s}}_\mathrm{act}}\xspace}
\newcommand{\Nactb}{\ensuremath{{N^{b}}_\mathrm{act}}\xspace}

\newcommand{\Nabos}{\ensuremath{{N^{s}}_{\uparrow}}\xspace}
\newcommand{\Nabob}{\ensuremath{{N^{b}}_{\uparrow}}\xspace}
\newcommand{\Nbels}{\ensuremath{{N^{s}}_{\downarrow}}\xspace}
\newcommand{\Nbelb}{\ensuremath{{N^{b}}_{\downarrow}}\xspace}
\newcommand{\Nobss}{\ensuremath{{N^{s}}_\mathrm{obs}}\xspace}
\newcommand{\Nobsb}{\ensuremath{{N^{b}}_\mathrm{obs}}\xspace}

\newcommand{\Snot}{\ensuremath{{\mathcal{S}_0}}\xspace}
\newcommand{\Sone}{\ensuremath{{\mathcal{S}_1}}\xspace}
\newcommand{\Bnot}{\ensuremath{{\mathcal{B}_0}}\xspace}
\newcommand{\Bone}{\ensuremath{{\mathcal{B}_1}}\xspace}

\newcommand{\poiss}{\ensuremath{\texttt{poiss}}\xspace}
\newcommand{\rhot}{\ensuremath{\rho_\mathrm{thr}}\xspace}
\newcommand{\w}{\ensuremath{w}\xspace}

\newcommand{\vd}{\ensuremath{D}\xspace}
\newcommand{\dl}{\ensuremath{\mathrm{d}_\mathrm{L}}\xspace}
\newcommand{\ud}{\mathrm{d}}
\newcommand{\md}{\ensuremath{\mathcal{D}}\xspace}
\newcommand{\dete}{\ensuremath{\rho_\uparrow}\xspace}
\newcommand{\xmax}{\ensuremath{x_\mathrm{max}}\xspace}
\newcommand{\mul}{\ensuremath{\mu_\lambda}\xspace}
\newcommand{\mut}{\ensuremath{\mu^T_\lambda}\xspace}

\newcommand{\sil}{\ensuremath{\sigma_\lambda}\xspace}

\newcommand{\Nsa}{\ensuremath{N_\mathrm{samples}}\xspace}

\newcommand{\dr}{\ensuremath{{S}}\xspace}
\newcommand{\db}{\ensuremath{{B}}\xspace}

\newcommand{\vr}{\ensuremath{\mathcal{R}}\xspace}

\newcommand{\nums}{\ensuremath{{N}^s}\xspace}
\newcommand{\numb}{\ensuremath{{N}^b}\xspace}
\newcommand{\N}{{\nums}\xspace}
\newcommand{\Ntri}{\ensuremath{N^{\mathrm{tr}}}\xspace}
\newcommand{\Nntri}{\ensuremath{N^{\mathrm{nt}}}\xspace}

\renewcommand{\D}{\ensuremath{D}\xspace}
\newcommand{\Dt}{\ensuremath{D^\mathrm{tr}}\xspace}

\newcommand{\refp}[1]{(\ref{#1})\xspace}

\def\blackbgd{0}

\newcommand{\LIGOlabMIT}{LIGO Laboratory, Massachusetts Institute of Technology, 185 Albany St, Cambridge, MA 02139, USA.}
\newcommand{\MKI}{Department of Physics and Kavli Institute for Astrophysics and Space Research, Massachusetts Institute of Technology, 77 Massachusetts Ave, Cambridge, MA 02139, USA.}
\newcommand{\StonyBrook}{Department of Physics and Astronomy, Stony Brook University, Stony Brook NY 11794, USA.}
\newcommand{\CCA}{Center for Computational Astrophysics, Flatiron Institute, New York NY 10010, USA.}
\newcommand{\bham}
{School of Physics and Astronomy \& Institute for Gravitational Wave Astronomy, University of Birmingham, Birmingham, B15 2TT, UK.}
\newcommand{\Vanderbilt}{Department of Physics and Astronomy, Vanderbilt University, 2301 Vanderbilt Place, Nashville, TN 37235, USA.}

\usepackage[section=section]{glossaries}

\newglossaryentry{Alpha}
{
  name=$\alpha(\vl)$,
  symbol=$\alpha(\vl)$,
  sort=Alpha,
  description={The fraction of physical sources that are detectable by our experiment, according to some detection threshold. It depends on the shape population hyper parameters \vl}
}

\newglossaryentry{Ns}
{
  name=\nums,
  symbol={\nums},
  sort=Ns,
  description={The total number of physical sources, detectable and not, produced in the universe. It is generally a function of time and hyper parameters \La}
}

\newglossaryentry{NsRateTime}
{
  name=\ensuremath{\dr_t},
  symbol=\ensuremath{\dr_t},
  sort=SnRateTime,
  description={The number of physical sources generated per unit time, i.e. \ensuremath{\dr_t=d\nums/dt}. It is generally a function of time and hyper parameters \La}
}

\newglossaryentry{NsRateVol}
{
  name=\ensuremath{\dr_{t,V_c}},
  symbol=\ensuremath{\dr_{t,V_c}},
  sort=SnRateVol,
  description={The number of physical sources generated per unit time per unit comoving volume, i.e. \ensuremath{\dr_{t,V_c}=\frac{\ud^2\nums}{\ud t \ud V_c}}. It is generally a function of time and hyper parameters \La}
}

\newglossaryentry{NsRateAll}
{
  name=\ensuremath{\dr_{\vec{\kappa}}},
  symbol=\ensuremath{\dr_{\vec{\kappa}}},
  sort=SnRateAll,
  description={The differential rate over a set of parameters belonging to a vector of $\vec\kappa$, i.e. $\dr_{\vec{\kappa}} \equiv \frac{\ud^n \nums}{\ud \kappa_1 \ldots \ud \kappa_n} $. It might be a function of time, hyper parameters \La and $\vec{\kappa}$}
}

\newglossaryentry{NaboveS}
{
  name=\Nabos,
  symbol={\Nabos},
  sort=Nsa,
  description={The number of physical sources that are above the detection threshold, and are thus detectable (though not necessarily detected!). It is generally a function of time and hyper parameters \La}
}

\newglossaryentry{Nsb}
{
  name=\Nbels,
  symbol={\Nbels},
  sort=Nsb,
  description={The number of physical sources that are below the detection threshold, and thus cannot be detected.  It is generally a function of time and hyper parameters \La}
}

\newglossaryentry{Nsex}
{
  name=\Nexps,
  symbol={\Nexps},
  sort=Nsex,
  description={The number of physical sources that are expected to be produced over some time interval. They might not all be detectable!  It is generally a function of time and hyper parameters \La}
}

\newglossaryentry{Nsac}
{
  name=\Nacts,
  symbol={\Nacts},
  sort=Nsac,
  description={The number of physical sources that were \emph{actually} produced over some time interval if $\Nexps$ were expected. These two numbers are not the same due to the Poissonian nature of the generation mechanism. They might not all be detectable!   It is generally a function of time and hyper parameters \La}
}

\newglossaryentry{Nsob}
{
  name=\Nobss,
  symbol={\Nobss},
  sort=Nsob,
  description={The number of physical sources that are actually observed, over the experiment time.  It is generally a function of time and hyper parameters \La}
}

\newglossaryentry{Nb}
{
  name=\numb,
  symbol={\numb},
  sort=Nb,
  description={The total number of background sources, detectable and not. It is generally a function of time}
}

\newglossaryentry{Nbex}
{
  name=\Nexpb,
  symbol={\Nexpb},
  sort=Nbex,
  description={The number of background events that are expected to be produced over some time interval. They might not all be detectable! It might be time dependent}
}

\newglossaryentry{Nbac}
{
  name=\Nactb,
  symbol={\Nactb},
  sort=Nbac,
  description={The number of background events that were \emph{actually} produced over some time interval if $\Nexpb$ were expected. These two numbers are not the same due to the Poissonian nature of the generation mechanism, which we assume in these Chapter. They might not all be detectable! It might be time dependent}
}

\newglossaryentry{NbRateTime}
{
  name=\ensuremath{\db_t},
  symbol=\ensuremath{\db_t},
  sort=BRateTime,
  description={The number of background sources generated per unit time, i.e. \ensuremath{\db_t=d\numb/dt}. It is generally a function of time}
}

\newglossaryentry{Nba}
{
  name=\Nabob,
  symbol={\Nabob},
  sort=Nba,
  description={The number of background sources that are above the detection threshold, and are thus detectable (though not necessarily detected!). It is generally a function of time}
}

\newglossaryentry{Nbb}
{
  name=\Nbelb,
  symbol={\Nbelb},
  sort=Nbb,
  description={The number of background sources that are below the detection threshold, and thus cannot be detected. It is generally a function of time}
}

\newglossaryentry{Nbob}
{
  name=\Nobsb,
  symbol={\Nobsb},
  sort=Nbob,
  description={The number of background sources that are actually observed, over the experiment time. It is generally a function of time}
}

\newglossaryentry{Ntri}
{
  name=\Ntri,
  symbol={\Ntri},
  sort=Ntri,
  description={The number of trigger that are actually observed over the experiment time. This might include both background sources and interesting physical sources.  It is generally a function of time and hyper parameters \La (through the sources)}
}

\newglossaryentry{Nsa}
{
  name=\Nsa,
  symbol={\Nsa},
  sort=Nsa,
  description={The number of samples that have been drawn from a distribution (usually a posterior density function or a likelihood). It might be different for different sources, though in these Chapter we assume it is always to same}
}

\newglossaryentry{thetavec}
{
  name=\vtheta,
  symbol={\vtheta},
  sort=thetavec,
  description={The vector of (usually unknown) parameters that characterize each and every source. Either us or someone else usually generates a finite number of fair draws from the posterior density function of \vt}
}

\newglossaryentry{Lambdavec}
{
  name=\La,
  symbol={\La},
  sort=Lambdavec,
  description={The vector of (usually unknown) hyper parameters that characterize the overall population. It might be further split in a subset of shape parameters and a scale parameter}
}

\newglossaryentry{lambdavec}
{
  name=\vl,
  symbol={\vl},
  sort=lambdavec,
  description={The vector of (usually unknown) shape hyper parameters that characterize the overall population. It can control e.g. the ratio of black holes generated at different masses}
}

\newglossaryentry{msun}
{
  name=\msun,
  symbol={\msun},
  sort=msun,
  description={The mass of our Sun today}
}

\newglossaryentry{pdet}
{
  name=\ensuremath{p(\dete| \vt)},
  symbol=\ensuremath{p(\dete| \vt)},
  sort=pdet,
  description={The probability that a trigger with true parameters \vt is detectable (which does not necessarily mean it is detected!)}
}

\newglossaryentry{DataRealiz}
{
  name=\md,
    symbol=\md,
  sort=DataRealiz,
  description={For a given trigger, the full set of data that could conceivably be produced by the experiment}
}

\newglossaryentry{DataDet}
{
  name=$\md_{\uparrow}$,
    symbol=$\md_{\uparrow}$,
  sort=DataDet,
  description={For a given trigger, the set of data that could conceivably be produced by the experiment and for which the trigger would be detectable}
}
\newglossaryentry{DataNodet}
{
  name=$\md_{\downarrow}$,
    symbol=$\md_{\downarrow}$,
  sort=DataNodet,
  description={For a given trigger, the set of data that could conceivably be produced by the experiment and for which the trigger would not be detectable}
}

\newglossaryentry{Data}
{
  name=\D,
    symbol=\D,
  sort=Data,
  description={The set of data \emph{actually} collected by the experiment. It usually is partitioned in a set that contains triggers, and another that does not contain triggers}
}

\newglossaryentry{DataA}
{
  name=\Dt,
    symbol=\Dt,
  sort=DataA,
  description={The set of data segments that contains triggers}
}

\newglossaryentry{DataI}
{
  name=$d_i$,
    symbol=$d_i$,
  sort=DatazI,
  description={A piece of data of small duration $\delta t$ that has been found to contain a trigger at time $t_i$}
}

\newglossaryentry{sort=DataJ}
{
  name=$\bar {d}_j$,
    symbol=$\bar {d}_j$,
  sort=DatazJ,
  description={A piece of data covering the time interval $[t_j,t_j+\delta t]$ that has been found not to contain any trigger}
}

\newglossaryentry{H}
{
  name={\maH},
    symbol=\maH,
  sort=H,
  description={Generically speaking, all information that is relevant to describe our experiment and data analysis. The hypothesis  \maH always go on the right side of conditional bars in our probability density functions, but we often omit it to simplify the notation}
}

\newglossaryentry{HL}
{
  name={\maHL},
    symbol=\maHL,
  sort=HL,
  description={All information relevant to the hyper analysis, e.g. the parametrization of model, the range of validity of the hyper parameters, the population priors, etc. It will be omitted unless its absence can create ambiguity}
}

\newglossaryentry{HP}
{
  name={\maHP},
    symbol=\maHP,
  sort=HP,
  description={All information relevant for the parameter estimation analysis of individual sources, e.g. the parametrization of model, the range of validity of the event parameters, the sampling prior used in MCMC, Nested sampling, or similar algorithm. In general the prior on the parameters used for sampling, $\pi(\vt | \maHP)$, will be different from what implied by the population model, $\pi(\vt | \La \maHL)$. It will be omitted unless its absence can create ambiguity}
}

\newglossaryentry{PrioPop}
{
  name=Population prior,
    symbol=PrioPop,
  sort=PopPrio,
  description={$\pi(\vt | \La\,\maHL)$. The expected distribution of the individual source parameters \vt given the population hyper parameters \La and the relative modeling \maHL. \maHL is often dropped unless needed}
}

\newglossaryentry{PrioPE}
{
  name=Sampling prior,
    symbol=PrioPE,
  sort=SamPrio,
  description={$\pi(\vt | \maHP)$. The expected distribution of the individual source parameters \vt used in an MCMC or similar algorithm while sampling the posterior of the individual triggers, given the settings used for the sampling \maHP}
}

\newglossaryentry{Source}
{
  name=Source,
    symbol=Source,
  sort=Source,
  description={A physically interesting event, that we are trying to detect and use for our inference}
}

\newglossaryentry{Background}
{
  name={Background},
    symbol=Background,
  sort=Background,
  description={An (usually uninteresting) background event that might be produced by any mechanism not related to the physics we are after. Background events can contaminate our sample}
}

\newglossaryentry{HypP}
{
  name={Hyper parameters},
    symbol=Hyper parameters,
  sort=Hyper parameters,
  description={The (usually unknown) set of parameters that control the properties of the individual sources in the population. We would like to infer the value of the hyper parameters, given a collection of detected sources. We use the symbol \La for the hyper parameters that fully characterize a model \maHL}
}

\newglossaryentry{Population}
{
  name={Population},
    symbol=Population,
  sort=Population,
  description={The true set of physical or astrophysical sources that we are trying to characterize by mean of a model described by some hyper parameters \La. In these notes we assume that only one astrophysical population exists that is producing all the sources. In real life there could be several (e.g. one for binary black holes and one for binary neutron stars), each described by its one set of hyper parameters and modeling \maH}
}

\newglossaryentry{rho}
{
  name={$\rho$},
    symbol=$\rho$,
  sort=rho,
  description={See SNR}
}

\newglossaryentry{rhodete}
{
  name={$\dete$},
    symbol=$\dete$,
  sort=rhodete,
  description={Indicates a signal-to-noise ratio above a predetermined detection threshold, i.e. $\rho>\rhot$, and hence indicates a detectable trigger}
}

\newglossaryentry{rhot}
{
  name={$\rhot$},
    symbol=$\rho$,
  sort=rho,
  description={The threshold value of the SNR. A source is detectable if $\rho>\rhot$}
}

\newglossaryentry{SNR}
{
  name={SNR},
    symbol=SNR,
  sort=SNR,
  description={The signal-to-noise ratio of an event. It depends on the true source parameters as well as on the properties of the detector. It can be used to decide if something is detectable}
}

\newglossaryentry{Trigger}
{
  name={Trigger},
    symbol=Trigger,
  sort=Trigger,
  description={A trigger recorder by our search. It is above the detection threshold, but we do not know if it is a real source or a background event}
}

\newglossaryentry{Tilda}
{
  name={$\sim$},
    symbol=$\sim$,
  sort=Tilda,
  description={Used to indicate a fair draw from a distribution. E.g. $x \sim p(x | \maH)$ means that $x$ is drawn from the distribution $p(x|\maH)$}
}

\newglossaryentry{T}
{
  name={$T$},
    symbol=$T$,
  sort=T,
  description={The duration of the experiment. When we talk of the number of triggers that have occurred, or are detected, or detectable, we might implicitly mean ``over the duration of the experiment $T$''}
}

\newglossaryentry{Occurrence}
{
  name={Occurrence},
    symbol=Occurrence,
  sort=Occurrence,
  description={The occurrence of a trigger just implies it happened, not necessarily that it was detected (or detectable). Everything that is detected or detectable occurred, but not everything that has occurred is detected or detectable}
}

\newglossaryentry{Detectable}
{
  name={Detectable},
    symbol=Detectable,
  sort=Detectable,
  description={A trigger is detectable if it would have a detection statistics above our pre-determined threshold. Detectable does not necessarily implies detected (e.g. the instrument might not be taking data at that time). Everything that is detected was detectable, but not everything that was detectable is detected}
}

\newglossaryentry{Detected}
{
  name={Detected},
    symbol=Detected,
  sort=Detected,
  description={A trigger is detected if it is above the detection threshold and we recorded the data that contains it. This is what we use for our inference. Everything that is detected was detectable and has occurred}
}

\newglossaryentry{Not}
{
  name={$\neg$},
    symbol=$\neg$,
  sort=Not,
  description={Logical negation of the proposition that follows it. $\neg A$ is true if $A$ is false and it is false if $A$ is true}
}

\newglossaryentry{And}
{
  name={$\land$},
    symbol=$\land$,
  sort=And,
  description={Logical ``and'' of two  propositions. $A\land B$ is true if and only if both $A$ and $B$ are true}
}

\makeglossaries

\if\blackbgd1
\pagecolor{black}
\color{white}
\else
\fi

\begin{document}
\title*{Inferring the properties of a population of compact binaries in presence of selection effects}
\titlerunning{Inferring populations with selection effects} 
\author{\large Salvatore Vitale$^*$, Davide Gerosa, Will M.~Farr, Stephen~R.~Taylor}
\authorrunning{Vitale et al.}
\institute{
Salvatore Vitale \at \MKI  \at \LIGOlabMIT  \at 
$^*$ To whom correspondence should be addressed:  \href{mailto:svitale@mit.edu}{svitale@mit.edu}
\and Davide Gerosa \at \bham \at \href{mailto:d.gerosa@bham.ac.uk}{d.gerosa@bham.ac.uk}
\and Will M. Farr \at \StonyBrook \at \CCA \at \href{mailto:will.farr@stonybrook.edu}{will.farr@stonybrook.edu}
\and Stephen~R.~Taylor \at \Vanderbilt \at \href{mailto:stephen.r.taylor@vanderbilt.edu}{stephen.r.taylor@vanderbilt.edu}
}

\maketitle

\vspace{-2cm}
\abstract{
Shortly after a new class of objects is discovered, the attention shifts from the properties of the individual sources to the question of their origin: do all sources come from the same underlying population, or several populations are required? What are the properties of these populations? 
As the detection of gravitational waves is becoming routine and the size of the event catalog increases, finer and finer details of the astrophysical distribution of compact binaries are now within our grasp. This Chapter presents a pedagogical introduction to the main statistical tool required for these analyses: hierarchical Bayesian inference in the presence of selection effects. All key equations are obtained from first principles, followed by two examples of increasing complexity. Although many remarks made in this Chapter refer to gravitational-wave astronomy, the write-up is generic enough to be useful to researchers and graduate students from other fields.
}

\vspace{0.5cm}

\noindent\textbf{Keywords}  Bayesian inference, source populations, selection effects, gravitational waves, black holes.

\section{Introduction}

Discovery can happen in several stages. First, one detects an unexpected or long sought-after signal. This source, and those that follow, each have their own distinct properties that make them unique, prompting speculation of their respective origin. As the number of sources increases, however, the attention moves from the properties of individual sources to using the ensemble of sources to learn something about the properties of the underlying population. Taken together, the signals 
might tell us something about the \emph{common} physical, astrophysical, biophysical, $\ldots$, cosmological processes that are generating \emph{all} the sources. The analysis can be made complicated by the fact that not all sources are equally easy to detect~\cite{1922MeLuF.100....1M,1925MeLuF.106....1M}, due to the sensitivity of the detector or other constraints. If not properly accounted for these selection effects will result in biased inference of the population that will usually grow with the population size~\cite{Mandel:2018mve}. This is called a selection (or Malmquist) bias.

The set of sources can really be anything, since the high-level techniques are agnostic to the experiment and messenger; domain specificity is encapsulated in the physics describing the detector selection effects and the mechanisms giving rise to the source population. A very incomplete and astronomy-biassed list of fields within which population inference has been employed includes neutrinos~\cite{Loredo:2001rx}, exoplanets~\cite{Foreman-Mackey:2014,2015ARA&A..53..409W}, gamma ray bursts~\cite{1995ApJS...96..261L,Loredo:1997qf}; or even heterogeneous sets of different messengers~\cite{Biscoveanu:2019bpy,Hayes:2019hso}. Although this chapter focuses mostly on transient gravitational-wave (GWs) signals emitted by compact binary coalescences~\cite{LIGOScientific:2018jsj,LIGOScientific:2018mvr}, it is generic enough to be useful for many scientists dealing with population analysis in the presence of selection effects. Examples in GW astronomy include inference on the stellar progenitors of stellar-mass merging black-holes~\cite{LIGOScientific:2018jsj,2018PhRvD..98h3017T,2019MNRAS.484.4216R,Talbot:2018cva,Zevin:2017evb,Vitale:2015tea,Farr:2017uvj,Talbot:2017yur,Fishbach:2019bbm}, the equation of state of neutron stars~\cite{2013PhRvL.111g1101D,Chatziioannou:2020pqz,Landry:2020vaw,Wysocki:2020myz}, the measurement of the Hubble constant~\cite{2017Natur.551...85A,2020PhRvD.101l2001G,2018Natur.562..545C,2012PhRvD..85b3535T,Chen:2020dyt,Mortlock:2018azx,Feeney:2018mkj}, and the origin of stochastic gravitational-wave background~\cite{2017PhRvL.118r1102T,2020MNRAS.496.3281S,2017MNRAS.468..404C}.

In this chapter you will not find any new scientific results, or new tricks that have not been suggested and exploited elsewhere. However, all relevant material has been collected here into a complete and consistent form that is ideal for new students learning the techniques for the first time, and hopefully also constitutes a helpful reference to seasoned experts.  At a pedagogical level, no steps in the derivations have been skipped, and the final expressions have been derived from first principles. This chapter might also be useful if you want to learn further details about the physics and statistics underlying much of the gravitational-wave literature, or really any branch of science dealing with similar issues of ensemble inference. This is certainly not the only available monograph that focuses on hierarchical population analysis with a pedagogical focus: the interested reader should also be sure to check out Refs.~\cite{Loredo:2001rx,Mandel:2018mve,2019PASA...36...10T,2019arXiv191112337L}.

The rest of this chapter is organized as follows: in Sec.~\ref{Sec.Setup} we present the basics of populations, sources, and hyper-parameters; in Sec.~\ref{Sec.Selection} we introduce selection effects and show how to calculate the fraction of detectable sources; in Sec.~\ref{Sec.HyperLike} and Sec.~\ref{Sec.HyperPost} we calculate the likelihood and the posterior density function of the population hyper-parameters, respectively;  in Sec.~\ref{Sec.OneDExample} we go through a simple one dimensional example, while a more complicated example inspired by gravitational-wave detections is given in Sec.~\ref	{Sec.GWExample}. A useful glossary, including a list of all the main symbols used in this chapter, is provided at the end.

\section{Source production, rates, populations}
 \label{Sec.Setup}
There is a physical process that generates sources at a rate
\beq
\dr_{t}(t)\equiv \frac{\ud \nums(t)}{\ud t}\,,
\label{Eq.Roft}
\eeq
where $\N(t)$ is the number of sources at time $t$. Throughout this chapter, we will use an index or letter ``s'' (possibly capitalized) to refer to \textbf{source} counts, rates, etc. When we need to refer to \textbf{background} events, we will instead use a letter or index ``b''.

If the source rate is constant in time one can write:
\beq
\frac{\ud \nums(t)}{\ud t}= \dr_t \implies \nums(t) = (t-t_0)\; \dr_t  +\nums(t_0)
\eeq
with $t_0$ an arbitrary reference time.

It often is useful to consider different types of differential source rates. For example, when studying the mergers of compact binaries through gravitational-wave emission~\cite{GW150914-DETECTION}, one is usually interested in the number of mergers, per unit time per unit comoving volume $V_c$~\cite{2010CQGra..27q3001A,2013ApJ...779...72D,GW150914-RATES,LIGOScientific:2018jsj}. To keep the notation compact and generalizable, we will just add more symbols to the underscore list:
\beq
\frac{\ud^2 \nums(t)}{\ud t\ud V_c}\equiv \dr_{t,V_c}(t)\,.
\eeq
More generally, if we  need to refer to the differential source rate over a set of parameters belonging to a vector $\vec\kappa$ we will write:

\beq
\dr_{\kappa_1,\ldots, \kappa_n}(t)\equiv \dr_{\vec{\kappa}}(t) \equiv \frac{\ud^n \nums(t)}{\ud \kappa_1 \ldots \ud \kappa_n} \equiv \frac{\ud \nums(t)}{\ud \vec{\kappa}} \,.
\label{Eq.RofKappa}
\eeq
Hereafter, we will not explicitly report the time dependence in every equation, but only in those where more than one time variable appear.

If each of the sources can be characterized by a set of parameters \vtheta (e.g. their masses, distances, etc)
we can talk of the \emph{fraction} of the \nums sources with parameters in the range $[\vtheta,\vtheta+\ud\vtheta]$:
\beq
p(\vtheta) = \frac{1}{\N} \frac{\ud \N}{\ud \vt} \equiv  \frac{1}{\N} \dr_{\vt}\,,
\label{Eq.pTheta}
\eeq
so that
\beq
\delta\N(\vt) \equiv \N p(\vt)\delta\vtheta
\label{Eq.dNofdTheta}
\eeq
is the \emph{number} of events with parameters  in the range $[\vtheta,\vtheta+\delta\vtheta]$.
These quantities are normalized as one would expect:
\beq
\int \ud \vtheta \;  \dr_{\vt} =\N
\eeq
and
\beq
\int \ud \vtheta \;p(\vtheta) = 1\,.
\eeq

Now, in general there will exist some underlying factors ---for example, the average metallicity at some redshift--- that can affect both the total number of sources, or their time rate,  and the distribution of the individual source parameters  \vt. We will call \vl the set of parameters that control the \emph{shape} of the distribution of the source parameters, and treat the overall source number \N as a free scale parameter. When necessary, we can refer to the shape and scale parameters together as:
\beq
\La \equiv \{\vl,\N\} \,.
\eeq
These are often called {\it hyper parameters}, to contrast them with the individual source parameters \vt.  The distribution of events' parameters thus becomes a function of \vl:
\beq
p(\vtheta)\rightarrow  \pi(\vtheta | \vl) \,.
\label{Eq.PopPrior1}
\eeq
We will refer to $\pi(\vt |\vl)$ as the \emph{population prior}, and use the symbol $\pi$ instead of $p$ to indicate that it is a prior distribution (in a Bayesian sense), i.e. a distribution that conveyes what we believe we know about \vt \emph{before} we analyze the data. Note that a prior distribution might be informed by \emph{previously analyzed} data: what data this distribution is prior to should be evident from the context.
In some of the literature, the population prior is indicated with an explicit ``pop'' label, but we prefer to avoid it since the presence of \vl on the right side of the conditional bar makes this unambiguous.

This makes most of the interesting quantities depend on the hyper parameters. For example the the number of sources per unit \vt, per unit time, and per unit time-\vt are:
\beq
 \dr_{\vt}(\La)= \N \pi(\vtheta|\vl) \,;\;\; \dr_t(\La) =  \frac{\ud \N(\La)}{\ud t}\,;\;\; \dr_{t\,,\vt}(\La) =  \frac{\ud^2 \N(\La)}{\ud t \ud \vt}\,.
\label{Eq.dNdTheta}
\eeq
In general we will not systematically report the dependence on \La, \N or \vl explicitly, unless needed, as it would only make formulae cumbersome. If in doubt, you can always refer to the Glossary, where the dependence of various key variables is explicitly spelled out.

One talks of hierarchical inference when a collection of detections from the population is used to infer $\La$.\footnote{In these article we only consider two layers of inference, but the procedure can in principle be generalized. One case where one could conceivably add third layer is the optimization of the kernel parameters of an algorithm that that infers hyperparameters.} %
The word {\it hierarchical} refers to the two levels of inference involved: one must first characterize the individual sources that are detected, i.e. measure \vt for all sources, then use them all to learn something about the population parameters  \La.

\section{Selection effects}\label{Sec.Selection}

In general, these observations will represent a biased sample of the true underlying population since some sources will be harder, or impossible, to detect.
Therefore, an important question we need to address is the following: given a source with a specific set of parameters \vt, would we be able to detect it, or not?  Usually, the answer is not a binary yes or not, because our detector is noisy, and depending the noise realization the same source might be more or less easy to detect. The best we can do is to fold into the analysis the fact that the noise is not entirely predictable, and calculate the \emph{probability} that a source with parameters \vt is \emph{detectable}: 

\beq
p(\mathrm{detectable} | \vt) = \int_{\md}  \ud \md \;p(\mathrm{detectable} | \md \vt) p(\md|\vt)
\label{Eq.PDetDef}
\eeq
where \md is the set of all possible data, i.e. the possible noise realizations,\footnote{That is, \md is \emph{not} the data that was actually collected, but rather all the data that might possibly have been collected.
} and we have used the sum rule of probability, since \md represents a complete set of mutually exclusive outcomes.
 We will assume that the analyst can calculate for each piece of data a \emph{detection statistics} $\rho_\md$ (e.g. a signal-to-noise ratio, a false-alarm rate probability, etc) which depends only on the data. Furthermore, we will assume that such a detection statistic can be used to establish if a piece of data contains a trigger, depending on how the detection statistics for that chunk of data compares to a pre-determined threshold.

In this case, we can split the integration domain into two disjoint sets: one corresponding to the data that will yield a detection statistic $\rho_\md$  above some pre-determined threshold:
\beq
\md_{\uparrow} \equiv \md : \rho_\md \ge\rho_\mathrm{thr}
\eeq
and its complement
\beq
\md_{\downarrow} \equiv \md : \rho_\md <\rho_\mathrm{thr}\,,
\eeq
where $\rho_\mathrm{thr}$ is the threshold detection statistic. Here and in the rest of these paper we will use an upward arrow, as in $ \md_{\uparrow}$, to refer to quantities that are associated with \emph{detectable} sources. Similarly, a downward arrow is used for quantities associated with undetectable sources. Following this syntax, we will use $\dete$ to indicated a trigger with SNR above the detection threshold; for example $p(\mathrm{detectable} | \vt)$ will just written as $p(\rho_\uparrow | \vt)$.
In some of the literature the label ``det'' is used to refer to sources which are detectable. We prefer the arrow because ``det'' could refer to both detectable and detected. These are very different concepts, which we want  to keep well separated.  Everything that is detected was detectable, but the opposite is not necessarily true (e.g. a detector might only be online 50\% of the time, which would halve the number of detected signals over some time range, but leave the number of detectable source unchanged).

Having partitioned the data this way, we can write:

\beqa
p(\dete | \vt) &=& \int_{\md} \ud \md \;p(\dete | \md \vt) p(\md|\vt) =  \notag\\
&=&  \int_{\md_{\uparrow}} \ud \md\; p(\dete | \md \vt) p(\md|\vt) +\int_{\md_{\downarrow}} \ud \md \;p(\dete | \md \vt) p(\md|\vt)
\eeqa
By construction, $p(\dete | \md \vt)=1$ [$p(\dete | \md \vt)=0$] if $\md\in\md_{\uparrow}$ [$\md \in \md_{\downarrow}$] thus the second integral is zero and one has:
\boxit{p(\dete | \vt) =  \int_{\md_{\uparrow}} \ud \md \;p(\md|\vt)\,.}{PDet}
This expression gives us the probability that a source with parameters \vt is detectable. We often need the total number of sources that are detectable by our instruments over some period of time, irrespective of their parameters. This can be done by calculating the number of sources produced per unit \vt, multiply by the probability of detecting those sources, and integrating over all possible \vt. This gives for the number of \emph{detectable} sources, given some population hyper parameters:
\beq\label{Eq.Ndet}
\Nabos \equiv  \int \ud\vt\; \frac{\ud \N}{\ud \vt}\, p(\dete|\vt)=  \N \int \ud\vt\; \pi(\vt |\lambda) p(\dete|\vt) \equiv \N \;\alpha(\vl)
\eeq
where we have used Eq.~\refp{Eq.pTheta} and defined the fraction of detectable events:
\boxit{\alpha(\vl) \equiv \frac{\Nabos}{\N} = \int \ud\vt\; \pi(\vt |\lambda) p(\dete|\vt)\,.}{Alpha}

Notice that while the probability that a source with true parameters \vt is detectable does (obviously) depend on the parameters, the \emph{criterion} used to decide if a piece of data contains a signal or not is based on the data, not on the parameters (which we would not usually know, anyways). In the expressions above this is explicitly indicated by the fact that the partition of the integrals is made according to some function of the data, $\rho_\md$~\cite{Mandel:2018mve}.  In practice the calculation of the detection statistic by a pipeline and the statistical distribution of real data are sufficiently complicated that the integral above is not analytically tractable.  In these circumstances it can still be estimated by the Monte-Carlo method by generating synthetic signals, placing them into the real data stream, applying the actual search pipelines to recover them, and recording the resulting detection statistics \cite{Farr:2019}.

\section{Hierarchical likelihood}\label{Sec.HyperLike}

The main goal of the Bayesian analyst is usually to calculate a posterior distribution for the population hyper parameters:
\beq
p(\La | \D \maHL)
\label{Eq.HyperLikeZero}
\eeq
where $D$ here indicates all of the available data produced by our detector(s) and  \maHL is a proposition that represents \emph{all} of the assumptions, settings, modeling, population parameters and range, population priors, etc, used in our hyper analysis (when we need to refer to a \emph{generic} set of settings we will use \maH without any subscript). It is important to remember that something like \maH is always present on the right of the conditional bars, even when it is not made explicit.  Usually, scientific papers provide in the body of the paper details about their analysis (start and stop time of the data segments used, data processing tools, windowing, sensitivity of the detector, the detection criterium, how much coffee the analyst had in the morning, etc): all of these factors, and more, should be indicated explicitly.\footnote{But you can clearly see why it would be prohibitive to explicitly write them all on the right of a conditional bar in every equation.}
With this in mind, we will not write \maH down explicitly in each and every equation, but will still write it down occasionally, especially if nothing else is on the right hand side of the conditional bar, or when things might otherwise be ambiguous because several possible settings are at play. We  will get back to the role of \maH in a few pages, when we hit an apparent contradiction and obtain two numbers that look like they should be the same, but are not.
In Eq.~\refp{Eq.HyperLikeZero} we are using a different symbol for the data than we did in the previous section, since before we wanted to refer to all possible realizations of the noise that could theoretically be produced, while keeping the source the same. Here $D$ is the \emph{actual} realization of the data our experiment recorded. In the case of GW detectors, for example, $D$ would be a time series reporting the strain.

In this section we will closely follow the derivation provided by Loredo and Lamb in the context of hierarchical inference of neutrino observations~\cite{Loredo:2001rx}, making minor changes to make the notation uniform with ours. %

\subsection{Data partitioning}\label{SuSec.DataPartition}

Let us imagine dividing the whole data set into a large sequence of small chunks, or segments, each of duration \dlt. The actual duration of \dlt is not important, as long as it is small enough that the probability of having more than one trigger (see definition in the next paragraph, or in the Glossary) in a time \dlt is negligible. We will exploit this soon.

At this point it's also worth reminding ourselves that, as explained in the previous section, we have identified a detection criterium $\rho_D > \rho_\mathrm{thr}$ based on the data, that we can use to decide if any piece of data contains something more than noise, or not.
If we use that criterium in all the small chunks of $D$, there will be two types of  data chunks: those for which $\rho_D$ is above threshold, and the others.
Notice that finding that a piece of data contains something above threshold, as just defined, does not automatically imply that it contains an interesting astrophysical signal. Unfortunately, GW detectors suffer from background transient events, which might trigger the detection criteria, but are not what we sought (these transients are usually called glitches~\cite{GW150914-DETCHAR}).  From now on we will use \emph{detection} uniquely to refer to astrophysical signals and \emph{trigger} to anything that rises above the detection threshold, no matter of whether it's astrophysical or a background.

Following Ref.~\cite{Loredo:2001rx}, we use the index $i$ to label the data chunks that have been found to contain a trigger. Therefore, $d_i$ is the i-th data chunk with a trigger, and covers the time range $\dlt_i \equiv [t_i, t_i + \dlt]$. We will use the index $j$ to label data chunks without triggers.
Therefore \ndj is the piece of data included in the time interval $\dlt_j \equiv [t_j, t_j + \dlt]$, and it does not contain any trigger.

The collection of $\{d_i\},\{ \ndj\},\forall \;i,j$ represents a complete and non-overlapping partition of the dataset $D$.
We can write the hyper posterior using Bayes theorem as:
\beq
p(\La | D \maHL) = p(D | \La \maHL) \frac{\pi(\La |\maHL)}{p(D |\maHL)}
\label{Eq.HyperPostDef}
\eeq
As in the rest of these notes, we use $\pi$ for priors: $\pi(\La |\maHL)$  thus represents the probability density function for \La that we use before looking at the \emph{current} set of GW data. The function
$p(D|\maHL)$ indicates the evidence of the data marginalized over the hyper parameters (also called the hyper evidence)
\beq
p(D|\maHL)\equiv \int \ud\La  \;p(D|\La \maHL) \pi(\La |\maHL)
\eeq
and usually won't play a role unless we are trying to do model selection among different population models~\cite{LIGOScientific:2018jsj}. We will get back to the hyper evidence later, but for now let us focus on the likelihood of the data, $p(D|\La \maHL)$. If we can assume that the noise has an autocorrelation length much smaller than \dlt, then different data segments are statistically independent, and the likelihood factorizes:
\beq
 p(D | \La \maHL) = p(\{d_i\},\{ \ndj\} | \La \maHL) = \left[\prod_{i=1}^{\Ntri} p\left(d_{i} | \La \maHL\right)\right]  \left[\prod_{j=1}^{\Nntri} p\left(\bar{d}_{j} | \La \maHL \right)\right]
 \label{Eq.HyperLike}
 \eeq
where the labels $tr$ and $nt$ stand for ``trigger'' and ``no-trigger''.

To explicitly calculate the likelihood, we must make some assumption on the generation mechanism of both real sources and background events.
We will assume that the probability for a source or a glitch to happen in any given data chunk is independent of what happened in other data chunks. This is a reasonable assumption in most cases. Certainly it is a good assumption for GW detectors at their current sensitivity.\footnote{It might become less robust of an assumption when the sensitivity of the detectors is such that a significant number of events will be gravitationally lensed, which will make what happens in the chunk $d_i$ not totally unrelated to what happened other segments.
} It is probably not a great assumption for glitches in GW detectors, as they sometimes come in groups~\cite{Zevin:2016qwy} (e.g. due to elevated seismic activity, or a faulty connection), but we will for the moment still stick to this assumption, because it is simple and allows making all the important points.\footnote{If the background production relevant for your detector clearly violates these assumptions, and one knows how to model it, most of our results can be modified accordingly.}
Under these hypotheses, both the distribution of the number of sources and backgrounds is Poissonian.

If \Nexps sources are expected to occur on average in a given time interval, the distribution of the number of source actually generated, \Nacts, is

\beq
p(\Nacts | \Nexps)= \poiss(\Nacts | \Nexps) = \frac{e^{-\Nexps} \left(\Nexps\right)^{\Nacts}}{\Nacts!}\,,
\label{Eq.PoissonSource}
\eeq
where $\poiss(O|E)$ is the Poisson probability for $O$ events occurring given that $E$ were expected.
Please notice that we have not yet said anything about whether these \Nacts sources are all detectable or not, these are not necessarily all detections!

What is the expected number of sources in each of our time intervals \dlt?
This can be simply written as the integral of the time-rate over the relevant time interval
\beq
\Nexps= \int_{\dlt} \ud t\; \dr_{t}= \int_{\dlt} \ud t \frac{\ud \N}{\ud t}\,,
\eeq

Note that the production mechanism depends on the hyper parameters, and so does \Nexps. Let us write this down explicitly this once:
\beq
\Nexps(\La)= \int_{\dlt} \ud t \dr_{t}(\La)= \int_{\dlt} \ud t \frac{\ud \N(\La)}{\ud t}\,.
\eeq
As long as the time rate is not a strong function of time, and can be treated as approximately constant over the small time interval \dlt, we can simplify this expression as:
\boxit{\Nexps(\La)\simeq  \dr_t(\La)\; \dlt\,.}{NexpS}
We will assume that this is always the case and use this simplification hereafter.

There is a small catch here: usually we will be interested in the expected number of sources generated in some time-interval \emph{as measured by the sources clocks}.  When the sources are distributed at cosmological distances, or are moving relativistically, there will be conversions to be done between the differential rate per unit time in the rest frame of the source, and the one in the detector frame. We will do this explicitly later, see~\Cref{Sec.GWExample}.

Similar expressions hold for the distribution of the actual number of background events:
\beq
p(\Nactb | \Nexpb) =\poiss(\Nactb | \Nexpb) = \frac{e^{-\Nexpb} \left(\Nexpb\right)^{\Nactb}}{\Nactb!}\,,
\label{Eq.PoissonBack}
\eeq
where the expected number of background events over the time interval \dlt is
\beq
\Nexpb=\int_{\dlt} \ud t \frac{\ud \numb}{\ud t}\equiv  \int_{\dlt} \ud t \db_{t}\,.
\eeq

In these expression, $\db_t(t)$ is defined as the number of background events per unit time, which we assume can be measured experimentally, or predicted theoretically or empirically.  As for the source rate, as long as $\db(t)$ does not vary too quickly, we can write the expected number of background events in the data chunk as:
\boxit{\Nexpb\simeq \dlt \db_t\,.}{NexpB}
We will assume this is always the case.\footnote{In some cases,  $\db_t$ might actually be time-independent, in which cases this equation is exactly valid.}

\subsection{Calculating the likelihood of data without a trigger}

We now have everything we need to explicitly calculate Eq.~\refp{Eq.HyperLike}. Please, make yourself comfortable, be sure you have a tea or your favorite drink nearby: this will take a while. We start by looking at a generic data chunk \ndj  where no triggers were found. We need to calculate $p(\ndj | \La)$.
This term is the probability of not reporting any trigger in the data chunk, given that the true value of the hyper parameters is \La, plus all other modeling and assumptions we have made.\footnote{Remember we are dropping \maHL on the right of the conditional bar (most of the times, at least).}

There are multiple reasons why no trigger might be found in the data: maybe nothing actually happened in that time interval and thus there was nothing to detect in the first place; or maybe a source or a background event \emph{did} occur, but their characteristics were such that the data did not match our criteria to claim a detection.

Let's formalize this mathematically.
Following ~\cite{Loredo:2001rx}, we will introduce the following propositions:
\beqa
\Snot &\equiv& \text{ ``No astrophysical source occurred in the time interval of interest''}\nonumber \\
\Sone &\equiv& \text{ ``Exactly one astrophysical  source occurred in the time interval of interest''}\nonumber\\
\Bnot &\equiv& \text{ ``No background event occurred in the time interval of interest''}\nonumber\\
\Bone &\equiv& \text{ ``Exactly one background event occurred  in the time interval of interest''}\nonumber
\eeqa
These prepositions are agnostic about the parameters of the trigger. Please notice the use of the word ``occurrence'', which does not imply either detectability or  detection (See the Glossary at the end of this document).

We can write the likelihood for non-detection by using the sum rule and the propositions above:
\beq
p(\ndj | \La) = p(\ndj (\Snot\Bnot + \Snot\Bone + \Sone\Bnot) | \La)\,.
\eeq
For us to be able to write this an an identity, it should be  the case that $\Snot\Bnot + \Snot\Bone + \Sone\Bnot=1$, while theoretically more than one trigger could occur during the time \dlt. However, because we picked \dlt to be very small, the probability of more than one trigger occurring in that time is so small that we treat it as zero.
We therefore can use the sum rule considering only the tree terms above, and expand as
\beq
p(\ndj | \La) =
 p(\ndj \Snot\Bnot |\La) + p(\ndj  \Snot\Bone | \La)  + p(\ndj \Sone\Bnot  | \La)\,,\label{Eq.PNoDetDef}
\eeq
where we used the fact that all of the four propositions above are mutually exclusive, within our analysis setup.

Let us now proceed and calculate each of these three terms.

\subsubsection{The good: $p(\ndj \Snot\Bnot |\La) $}

The term  $p(\ndj \Snot\Bnot |\La) $ represents the probability of no trigger being found in the data \emph{and} no signals occurring \emph{and} no background occurring, given the hyper parameters.

We can use the product rule and write it as:

\beq
p(\ndj \Snot\Bnot |\La) =p(\ndj | \Snot\Bnot  \La) p(\Snot\Bnot |\La) =p(\ndj | \Snot\Bnot  \La) p(\Snot | \La) p(\Bnot|\maH)\,,
\label{Eq.TheGood}
\eeq
where in the last step we have used the fact that \Snot and \Bnot are independent to factorize their joint probability, and the fact that \Bnot does not depend on the astrophysical source hyper-parameters to write $p(\Bnot|\La)=p(\Bnot|\maH)$.\footnote{We are adding a generic \maH on the right of the conditional bar, because we are horrified to see nothing there.}

The first term, $p(\ndj | \Snot\Bnot  \La)$, is trivial: the probability of nothing raising above the detection threshold given that nothing happened is 100\%:
\beq
p(\ndj | \Snot\Bnot  \La) =1\,.
\eeq
Notice how \La is not really playing a role here, since no source occurred, and it could have been removed from the right hand side of the conditional bar.

Let us now move to $p(\Bnot|\maH)$, that is the probability that no background occurred in the j-th time chunk. Using the Poisson distribution, Eq.~\refp{Eq.PoissonBack}, this is the probability of 0 occurrences given that $\db_t\dlt$ were expected:
\beq
p(\Bnot| \maH)= \poiss(0 | \Nexpb=\db_t \dlt) = e^{-\db_t \dlt}\,.
\label{Eq.PBnot}
\eeq
The same to $p(\Snot | \La)$ :
\beq
p(\Snot | \La) = \poiss (0 | \Nexps=\dr_t \dlt) = e^{- \dr_t \dlt}\,,
\label{Eq.PSnot}
\eeq
such that, putting all together, we obtain:
\boxit{p(\ndj \Snot\Bnot |\La) = e^{-\left(\db_t+\dr_t\right) \dlt}\,.}{TheGoodFinal}

\subsubsection{The bad: $p(\ndj  \Sone\Bnot | \La)$}\label{SuSec:TheBad}

Let's use the product rule on this term but without fully expanding all terms:
\beq
p(\ndj  \Sone\Bnot  | \La) =p(\ndj \Sone | \Bnot  \La) p(\Bnot | \La) = p(\ndj  \Sone | \La)p(\Bnot |\maH)\,.
\eeq

The very last term, i.e. the probability of no background event occurring, has already  been calculated in the previous subsection, Eq.~\refp{Eq.PBnot}:
\beq
p(\Bnot |   \maH)= e^{-B_t \dlt}\,.
\eeq

Let us now look at the first term, the probability that one source occurs \emph{and} no triggers are reported, given \La.
The proposition \Sone does not specify the parameters of the source that occurred, but in order to do explicit calculations we need to specify some value of parameters. This can be achieved by marginalizing over all of the possible source parameters:

\beq
p(\ndj  \Sone | \La) = \int \ud \vt p(\ndj  \Sone \vt | \La)\,,
\eeq
and using the product rule:
\beq
p(\ndj  \Sone | \La) = \int \ud \vt p(\ndj  \Sone \vt | \La) =  \int \ud \vt p(\ndj   |\Sone \vt \La) p(\Sone \vt | \La).
\eeq

Here, $p(\Sone \vt | \La)$ is the probability \emph{density} that one and only one event occurred, and that it had parameters within a small interval $\delta \vt$ around \vt. The probability itself is given by
\beq
p(\Sone \vt | \La) \delta\vt \,.
\label{Eq.RefPSoneVt}
\eeq
We need to be careful when calculating this: this is \emph{not} just the Poisson probability for the occurrence of one source with parameters \vt when $\Nexps(\vt,\La)$ sources with parameters \vt were expected. This would be:
\beqa
p(\text{One source occurrence with parameters \vt} | \La) =
 \poiss(1 | \Nexps(\vt,\La))\nonumber\\
=  \Nexps(\vt,\La) e^{- \Nexps(\vt,\La)} = (\dlt \delta\vt \dr_{t,\vt}) e^{- (\dlt \delta\vt \dr_{t,\vt}) }\,,
\label{Eq.OneSource}
\eeqa
where we have used the syntax of Eq.~\refp{Eq.RofKappa} to write the number of events in an interval \dlt and $\delta\vt$ as
\beq
\Nexps(\vt,\La) = \int_{\delta{\vt}} \int_{\dlt} \frac{\ud^2 \N}{\ud \vt \ud t}(\vt,\La) \ud t \ud\vt \simeq \frac{\ud^2 \N}{\ud \vt \ud t}(\vt,\La) \dlt\delta{\vt} \equiv \dlt \delta\vt \dr_{t,\vt}(\vt,\La)\,.
 \eeq

In fact Eq.~\refp{Eq.RefPSoneVt} expresses a much stronger requirement: namely that one event with parameters in the interval $\delta \vt$ around \vt occurred \emph{and that no other events with any other values of the parameters occurred}.
The simple Poisson probability we just calculated does not say anything about what else might have happened in that time interval, thus we need to explicitly calculate an extra term given by
\beq
p(\text{No source occurrence with any parameters different from \vt} | \La)\,.
\label{Eq.NoOtherSource}
\eeq
This looks like a Poisson probability for zero occurrences, but what is the number of expected events with parameters other than \vt? That is going to be the total number of expected events minus the number of expected events with parameters \vt~\footnote{The symbol $\neg$ is usually used to indicate the logical negation of the propositions that follows, e.g. $\neg A$ stands for ``not A''. We are committing here a slight abuse of notation and use $\neg \vt$ to indicate all values of the parameters outside of the  interval $\delta \vt$ around \vt.%
}:
\beq
\Nexps(\neg \;\vt) = \dr_t \dlt - \dr_{t,\vt}\; \dlt \delta\vt\,,
\eeq
such that Eq.~\refp{Eq.NoOtherSource} becomes:
\begin{align}
&p({\rm No source occurrence with any parameters different from \vt} | \La)
\nonumber \\
&= \poiss(0| \Nexps(\neg \;\vt))
\end{align}

The quantity we are calculating, Eq.~\refp{Eq.RefPSoneVt} is thus the logical ``and'' of the propositions expressed in Eq.~\refp{Eq.OneSource} and Eq.~\refp{Eq.NoOtherSource}:

\begin{align}
p(\Sone &\vt | \La)\delta\vt
= p(\text{One source occurrence with parameters \vt} | \La) \notag \\
&\times p(\text{No source occurrence with any parameters different from \vt} | \La)
\nonumber \\
&=  (\dlt \delta\vt \dr_{t,\vt}) e^{- \dlt \delta\vt \dr_{t,\vt} } e^{-\Nexps(\neg \;\vt)}  =  (\dlt \delta\vt \dr_{t,\vt}) e^{- \dlt \delta\vt \dr_{t,\vt} } e^{-\left( \dr_t \dlt - \dr_{t,\vt}\; \dlt \delta\vt \right)} 
\nonumber \\
&= \dlt \delta\vt \dr_{t,\vt}\; e^{- \dr_t \dlt}\,,
\end{align}
so that
\beq
p(\Sone \vt | \La)=\dlt \dr_{t,\vt}\; e^{- \dr_t \dlt}\,.
\label{Eq.PSoneVt}
\eeq
Notice that the term in the exponent is the expected number of event occurrences in the time interval, irrespectively of their parameters!

Finally, we need to calculate $p(\ndj   |\Sone \vt \La)$, this is the probability of recording a piece of data without any trigger above the threshold, given that one source occurred with parameters in the interval \intvt. This is one minus the probability of detecting
 the trigger, i.e.:
\beq
p(\ndj   |\Sone \vt \La) = 1 - p(\text{The trigger is detected}   |\Sone \vt \La)
\eeq
Notice that  we can drop \Sone and \La on the right of the conditional bar of the last term, since \vt already tell us all we need:
\begin{align}
p(\text{The trigger is detected}   |\Sone \vt \La) &=
p(\text{The trigger is detected}   |\vt )
\notag \\=
p(\text{The trigger is detectable}   |\vt )
&=
p(\dete|\vt)\,.
\end{align}
In the third line we have switched ``detected'' with ``detectable'' because in the calculation we are carrying out a piece of data ($\ndj$) \emph{was} collected. This excludes the possibility that a detectable source might have occurred but not being detected. Therefore, this is exactly the same quantity we defined earlier, in Eq.~\refp{Eq.PDetDef} and Eq.~\refp{Eq.PDet}

Putting all together yields
\boxit{p(\ndj  \Sone\Bnot | \La)= e^{-\left(B_t+\dr_t \right)\dlt} \dlt  \int \ud \vt \left[1- p(\dete|\vt)\right] \dr_{t,\vt}\,. }{TheBadFinal}

\subsubsection{The ugly: $p(\ndj  \Snot\Bone | \La)$}

The last term we need to compute is $p(\ndj  \Snot\Bone | \La)$. Let us use the product rule once more:
\beq
p(\ndj  \Snot\Bone | \La)=p(\ndj \Bone|  \Snot \La) p(\Snot |  \La) =p(\ndj \Bone | \maH) p(\Snot |  \La)\,,
\eeq
where we have used the fact that no source occurred, and hence we cannot learn anything about the data from \Snot and \La, to simplify one of the terms.

We already calculated $p(\Snot |   \La)$ in the previous subsections, Eq.~\refp{Eq.PSnot}:
\beq
p(\Snot |   \La)= e^{- \dr_t \dlt}\,.
\eeq
We now consider the other term, $p(\ndj \Bone | \maH)$. This is the probability that a background event occurred \emph{and} that no trigger was reported (\maH reminds us of the other things we assumed but are not explicitly listing: no sources occurred, our search algorithm was configured in a certain way and uses a $\rhot$ to select triggers, etc).
To make progress we need to follow a procedure similar to the one in Sec.~\ref{SuSec:TheBad}, and expand this across all the possible \vt:
\beq
p(\ndj \Bone | \maH)=\int \ud \vt p(\ndj \Bone \vt | \maH) =\int \ud \vt p(\ndj |\Bone \vt \maH) p(\Bone \vt  |\maH)\,.
\eeq
This might  seem puzzling at first: why are we using the astrophysical parameters of the sources, \vt, now that only background events are being considered? What we are doing is to \emph{assume} that background events, or at least the ones we care about, i.e. the ones that can be ``detected'', can be characterized with the same parameters that are used to describe the astrophysical sources. This is not entirely correct, and if we had a better model and parametrization for the background, we should use that instead.
However, If we think about it, this is not totally unreasonable either: since those background events are \emph{detected} by an algorithm that was looking for an astrophysical signal with some unknown set of parameters \vt, it is not too far fetched to assume that the background event too can be described by similar parameters. Obviously, in the case of a background, the meaning of the measured \vt is not immediately clear. \emph{Assuming} the background can be described with the same parameters as the sources, has the advantage of making the math easier, since we can literally follow the same steps of the previous subsection and calculate $p(\Bone \vt  |\maH)$ as the product of two terms: the probability of one background event occurring with parameters in \intvt \emph{and} the probability of zero background events happening with any other parameters. The result is:
\beq
p(\Bone \vt  |\maH) = \dlt B_{t,\vt}\; e^{- B_t \dlt}\,,
\eeq
where
\beq
B_{t,\vt} \equiv \frac{\ud^2  \numb}{\ud t \ud \vt}\,.
\label{Eq.BtVT}
\eeq
is the differential rate of background events per unit time and per unit \vt.

There is only one term left to calculate, $p(\ndj |\Bone \vt \maH)$ i.e. the probability that a background event occurring with parameters \vt would not be detectable. This is one minus the probability that it would be detectable:
\beq
p(\ndj |\Bone \vt \maH) = 1 - p(\dete |\Bone \vt \maH)\,.
\eeq
Here we are going to assume that our experiment and search algorithm are blind about the true nature of the trigger, whether it's an astrophysical source or a background event. If a background and a real source occurred with the same set of parameters \vt  in two pieces of data with the same noise realization, they will be both classified as a detection or a non-detection. This is a quite good assumption: if the detector or the algorithm \emph{had} a way of distinguish these two situations, we would use that as a tool to get rid of background events. The very fact that we are here talking about them implies that at least a sneaky fraction of the background events looks and walks like our real sources!

One gets:
\beq
p(\dete |\Bone \vt \maH)= p(\dete |\vt \maH)\,,
\eeq
which is \emph{the same} probability of detection we used in the previous section, cf. Eq.~\refp{Eq.PDetDef}. Putting all together we get:
\boxit{p(\ndj  \Snot\Bone | \La)= e^{- \left(\dr_t+\db_t\right) \dlt}  \dlt  \int{ \ud \vt  \left[1 - p(\dete | \vt )\right] \db_{t,\vt}}\,.}{TheUglyFinal}

\subsubsection{Putting the pieces together}\label{SuSec.Pieces}

We now have all the pieces we need to calculate Eq.~\refp{Eq.PNoDetDef}: %

\beqa
p(\ndj | \La) &=& p(\ndj \Snot\Bnot |\La) + p(\ndj  \Snot\Bone | \La)  + p(\ndj \Sone\Bnot  | \La)
\eeqa

From Eq.~\refp{Eq.TheGoodFinal}, Eq.~\refp{Eq.TheBadFinal}, and  Eq.~\refp{Eq.TheUglyFinal} we obtain:
\begin{align}
p(\ndj | \La) &= e^{-\left(\db_t+\dr_t\right) \dlt}\!+\!e^{-\left(\db_t+\dr_t \right)\dlt} \dlt  \int\! \ud \vt \left[1\!-\! p(\dete|\vt)\right] \dr_{t,\vt} 
\nonumber\\
&\qquad+ e^{- \left(\dr_t+\db_t\right) \dlt}  \dlt  \int{\! \ud \vt  \left[1 \!-\! p(\dete | \vt )\right] \db_{t,\vt}}
\nonumber\\
&= e^{-\left(\db_t+\dr_t\right) \dlt}  \left\{ 1+ \dlt \int \ud \vt \left[1- p(\dete|\vt)\right] \left(\db_{t,\vt}+ \dr_{t,\vt}\right) \right\}\,.
 \end{align}
We can simplify this expression considerably if we recognize that the second term in the curly brackets is much smaller than 1. Mathematically, this happens because we have taken $\dlt \ll1$, which implies $\dlt B_t \ll1$ and  $\dlt \dr_t \ll1$. Physically, this means that most of the times there are no triggers in our data.

We can exploit this by taking the logarithm of the non-detection probability:
\beq
\log p(\ndj | \La) \simeq - \dlt \left(\db_t+\dr_t\right)  +\dlt \int \ud \vt\left[ 1- p(\dete|\vt)\right] \left(\db_{t,\vt}+ \dr_{t,\vt}\right)
\,,
\label{Eq.ApproxLogNonDet}
\eeq
where we have used $\log(1+x)\simeq x$ for $x\ll1$.
The second term can be simplified as:

\beq
\dlt \int \ud \vt\left[ 1- p(\dete|\vt)\right] \left(\db_{t,\vt}+ \dr_{t,\vt}\right) = \dlt \left(\db_t+\dr_t\right)  - \dlt \int \ud \vt\; p(\dete|\vt) \left(\db_{t,\vt}+ \dr_{t,\vt}\right).
\eeq

Therefore, Eq.~\refp{Eq.ApproxLogNonDet} becomes:
\beq
\log p(\ndj | \La) \simeq - \dlt \int \ud \vt\; p(\dete|\vt) \left(\db_{t,\vt}+ \dr_{t,\vt}\right)\,,
\eeq
and switching back to the actual probability
\beq
p(\ndj | \La) \simeq e^{- \dlt \int \ud \vt\; p(\dete|\vt) \left(\db_{t,\vt}+ \dr_{t,\vt}\right)}\,.
\eeq

A last simplification can be made if the instrument noise is stationary, so that $p(\dete|\vt)$ is time independent:
\beqa
\int \ud \vt\; p(\dete|\vt)\dr_{t,\vt} &=&   \int \ud \vt\; p(\dete|\vt)\frac{\ud^2 \N}{\ud t \ud \vt}  =\frac{\ud}{\ud t} \int \ud \vt\; p(\dete|\vt)\frac{\ud \N}{\ud \vt} =   \frac{\ud \Nabos}{\ud t}
\eeqa
This is the time \emph{rate} of detectable astrophysical sources. One can similarly define a rate of detectable backgrounds and finally write:

\boxit{p(\ndj | \La) \simeq e^{- \dlt \left(\frac{\ud \Nabos}{\ud t}+\frac{\ud \Nabob}{\ud t}\right)}  }{PNonDetFinal}

This misleadingly simple expression, obtained with so much use of virtual ink, contains the selection effect of our instruments.

\subsection{Calculating the likelihood of data with a trigger}

Those of us with a significant attention span, will remember that we are only half-done, because we still need to calculate  the probability $p(\di | \La)$ of finding a trigger in the data chunk \di.

A trigger can be reported because a background event occurs, or because a source occurs. We can thus use the same propositions introduced in the previous section and write:

\beqa
p(\di | \La)&=&p(\di (\Snot\Bnot + \Snot\Bone + \Sone\Bnot) | \La) \nonumber\\
& =&  p(\di \Snot\Bnot |\La) + p(\di  \Snot\Bone | \La)  + p(\di \Sone\Bnot  | \La)\nonumber\\
& =&   p(\di  \Snot\Bone | \La)  + p(\di \Sone\Bnot  | \La) \label{Eq.PDetDef2}
\eeqa

One trivially has $p(\di \Snot\Bnot |\La)=0$ because we cannot possibly have recorded a trigger if nothing occurred. But what about noise fluctuations and false positive? Those \emph{are} background events, and thus would be captured by \Bone.

We therefore have two terms to calculate. Much of the heavy lifting was done in the previous section, thus this will be much faster.

\subsubsection{We got a source: $ p(\di \Sone\Bnot  | \La)$}

Just as we did in Sec.~\ref{SuSec:TheBad} we can use the product rule, and marginalize over all possible source parameters. We get:
\begin{align}
p(\di \Sone\Bnot  | \La) &= p(\Bnot |  \maH)p(\di \Sone | \La \Bnot) \notag\\&= p(\Bnot | \maH) \int \ud \vt p(\di \Sone \vt | \La ) 
\notag\\&=  p(\Bnot |  \maH) \int \ud \vt p(\di  | \Sone \vt \La ) p(\Sone \vt|\La)\,.
\end{align}

Both the $p(\Bnot| \maH)$ and $p(\Sone \vt|\La)$ have already been calculated, see Eq.~\refp{Eq.PBnot} and Eq.~\refp{Eq.PSoneVt}, respectively.
The remaining term $p(\di  | \Sone \vt \La )$ is the probability of getting the data \di given that a source occurred with parameters \vt. This is just the likelihood of the data:
\beq
p(\di  | \Sone \vt \La )= p(\di  | \vt )\,
\eeq
which is the very same function used to calculate $p(\dete | \vt)$ in Eq.~\refp{Eq.PDet}.

Putting it all together yields:
\boxit{p(\di \Sone\Bnot  | \La)= e^{-\left(\dr_t+ \db_t\right) \dlt} \dlt  \int \ud \vt p(\di  | \vt )  \dr_{t,\vt} \,. }{}
Notice that the integral here knows nothing about selection effects: the integration range must be performed across all possible \vt that the population allows for, regardless of whether they are detectable or not. Similarly, the likelihood function is the ``usual'' likelihood function, not truncated or affected in any way by the existence of selection effects.

\subsubsection{We got a background: $ p(\di \Bone\Snot  | \La)$}

We proceed exactly as in the previous subsection and obtain:
\boxit{p(\di \Bone\Snot  | \La)= e^{-\left(\dr_t+ \db_t\right) \dlt} \dlt  \int \ud \vt p(\di  | \vt )  B_{t,\vt}  \,.  }{}
Here again, we have assumed that the detector is blind to whether something with parameters \vt is a source or a background, hence obtaining the equalities:
\beq
p(\di  | \Sone \vt \La )= p(\di  | \Bone \vt \La )=p(\di  | \vt ) \,.
\eeq
\subsubsection{Putting the pieces together}

We can now add the terms obtained in the last two subsections and write Eq.~\refp{Eq.PDetDef2} as:

\beqa
p(\di | \La)=e^{-\left[\dr_t+ \db_t\right] \dlt}\dlt  \left[ \int \ud \vt\; p(\di  | \vt )  \left(\dr_{t,\vt}  + \db_{t,\vt} \right)    \right]\,.
\eeqa

\subsection{Finishing off}

We now have all we need to write down the posterior of \La, Eq.~\refp{Eq.HyperLike}. We are going to use the equations derived in the previous sections, explicitly reporting the times at which time-dependent quantities are evaluated:
\begin{align}
 &p(D | \La \maHL) = p(\{d_i\},\{ \ndj\} | \La \maHL) = \prod_{i=1}^{\Ntri} p\left(d_{i} | \La \maHL\right)\prod_{j=1}^{\Nntri} p\left(\bar{d}_{j} | \La \maHL \right)\nonumber\\
 &= \left\{\prod_{i=1}^{\Ntri}  e^{-\left[\dr_t(t_i)+ \db_t(t_i)\right] \dlt}\dlt  \int \ud \vt p(\di  | \vt )  \left[\dr_{t,\vt}(t_i)  + \db_{t,\vt}(t_i) \right]  \right\}  \nonumber\\
 &
\qquad\times  \left\{ \prod_{j=1}^{\Nntri} e^{- \dlt \left[\frac{\ud \Nabos}{\ud t}(t_j)+\frac{\ud \Nabob}{\ud t}(t_j)\right]}\right\}
 \end{align}
 We have not put any indices in \dlt since we have assumed the data was partitioned in intervals with the same duration (this condition can be relaxed quite trivially, if necessary).

We can write the product of exponentials as
\beq
\prod_{i=1}^{\Ntri}  e^{-\left[\dr_t(t_i)+ B_t(t_i)\right] \dlt}=e^{- \dlt  \sum_{i=1}^{\Ntri}\left[\dr_t(t_i)+ B_t(t_i)\right] } = e^{-\int_{T^\mathrm{tr}} \ud t\left[ \dr_t(t)+ B_t(t)\right]}
\label{Eq.ProdT}
\eeq
where we have defined $T^\mathrm{tr}= \Ntri \dlt $ as the total amount of time during which triggers were reported. Similarly for the segments without triggers:
\beq
\prod_{j=1}^{\Nntri} e^{- \dlt \left[\frac{\ud \Nabos}{\ud t}(t_j)+\frac{\ud \Nabob}{\ud t}(t_j)\right]}= e^{- \dlt \sum_{j=1}^{\Nntri} \left[\frac{\ud \Nabos}{\ud t}(t_j)+\frac{\ud \Nabob}{\ud t}(t_j)\right]} = e^{- \int_{T^\mathrm{nt}} \ud t  \left[\frac{\ud \Nabos}{\ud t}(t)+\frac{\ud \Nabob}{\ud t}(t)\right]}\label{Eq.ProdNT}
\eeq
where $T^\mathrm{nt}\equiv N^\mathrm{nt} \dlt$, and we note that in both this and the previous expression, the integral will cover disjoint data segments.

If the rate of background and astrophysical source are small enough that most of the times nothing is occurring, i.e.
\beq
T^\mathrm{tr} \ll T^\mathrm{nt}\,,
\eeq
we can neglect the exponential in Eq.~\refp{Eq.ProdT}, while keeping the integral over non-triggering times, Eq.~\refp{Eq.ProdNT}. Furthermore, in Eq.~\refp{Eq.ProdNT} we can also change the integration range from $T^\mathrm{nt}$ to the total experiment time $T$, since the two only differ by the total time with triggers, which is small.

Therefore, we finally obtain:

\begin{empheq}[box={\fboxsep=6pt\fbox}]{alignat=2}
 p(D | \La \maHL) &=&  e^{- \int_{T} \ud t  \left(\frac{\ud \Nabos}{\ud t}(t)+\frac{\ud \Nabob}{\ud t}(t)\right)} \prod_{i=1}^{\Ntri}  \dlt  \int \ud \vt p(\di  | \vt )  \left(\dr_{t,\vt}(t_i)  + B_{t,\vt}(t_i) \right)\,,\label{Eq.POfDFinal}
\end{empheq}
which is identical to Eq. A26 of Ref. \cite{Loredo:2001rx}, modulo a different syntax.

\subsection{Special cases}

The expression we found can be simplified a bit further if more assumptions can be made about the background or source rates.

\subsubsection{Constant background rate}

If the background rate is constant in time, then we have:
\beq
\int_{T} \ud t \db_t = T  \db_t = \numb\,.
\eeq
This is the total number of background events over the experiment time, i.e. the experiment time multiplied by the constant rate of background triggers. Similarly, the total number of \emph{detectable} background triggers over the duration of the experiment becomes
\beq
\int_{T}  \ud t \frac{\ud \Nabob}{\ud t} = T  \frac{\ud \Nabob}{\ud t} = \Nabob\,.
\eeq
Finally, (cfr Eq.~\refp{Eq.BtVT}) the differential background rate per unit \vt is now constant in time:
\beq
 B_{t,\vt}(t) =  B_{t,\vt} = \frac{B_{\vt}}{T}%
\eeq
and the likelihood simplifies as:
\beq
 p(D | \La \maHL) =  e^{-  \Nabob - \int_{T} \ud t  \frac{\ud \Nabos}{\ud t}(t)}   \prod_{i=1}^{\Ntri}  \dlt  \int \ud \vt p(\di  | \vt )  \left[\dr_{t,\vt}\,(t_i)  + \frac{B_{\vt}}{T} \right]
\eeq

\subsubsection{Constant background and source rate}

Let us now suppose that the rate of source occurrence is also constant in time. We have
\beq
\dr_t(t)=\dr_t = \frac{N^s}{T}\,,
\eeq
while the term in the exponential becomes:
\beq
\int_{T} \ud t \frac{\ud \Nabos}{\ud t} = T  \frac{\ud \Nabos}{\ud t} = \Nabos \,,
\eeq
which is the total number of detectable sources over the duration of the experiment.

The likelihood thus simplifies as:
\beq
 p(D | \La \maHL) =  e^{-  (\Nabob + \Nabos)}  \prod_{i=1}^{\Ntri} \frac{\delta t}{T} \int \ud \vt p(\di  | \vt )  \left(\dr_{\vt} + B_{\vt} \right)\,,
\eeq
where now the exponential contains the total number of expected triggers, irrespective of their nature. For each chunk with a trigger, we integrate the likelihood times the differential background and source rate per unit \vt, without any selection effect.

\subsubsection{Constant source rate and no background}

Finally, the most optimistic scenario, which is also the one that yields the most compact expression, is the one where we can assume both that the source rate is constant in time and that the background rate is zero.
This is a putative scenario where instruments and search algorithms are so accurate that we can safely exclude the possibility that our detection criteria can be met by anything else besides the occurrence of a real source.

In this ideal world, one has:
\boxit{p(D | \La \maHL) =  e^{-  \Nabos} \prod_{i=1}^{\Ntri}  \frac{\delta t}{T}\int \ud \vt p(\di  | \vt ) \frac{\ud N^s}{\ud \vt}(\La)\,.%
}{LoredLamb}

This is the fundamental equation for hierarchical inference used in most of the GW literature.
Please notice that we will be using \Ntri for the product even if we are assuming that there is no background, thus all triggers are sources. This is because we want to avoid any possible confusion with the true total number of sources (\N). This is fully consistent with our definition of trigger, see the Glossary at the end of this document.

\section{Hierarchical posterior}\label{Sec.HyperPost}

\subsection{The full hyper posterior}

We now derive the posterior on the hyper parameters \La. We will work with the simplest case, with no background and constant source rate (see \cite{2019MNRAS.484.4008G} for an analysis that accounts for background). In this case we can elaborate Eq.~\refp{Eq.LoredLamb} as:\footnote{We have removed the term $ {\delta t}/{T}$ from inside the product, since it's a constant term that does not depend on \La, and thus will not affect its inference.}
\beq
p(D | \vl \N) = e^{-  \Nabos}  \prod_{i=1}^{\Ntri}   \int \ud\vt\; \frac{\ud \N}{\ud \vt}(\vl)\, p(d_i |\vt) = e^{-  \Nabos}  \prod_{i=1}^{\Ntri} \N  \int \ud\vt\; \pi(\vt|\vl)\, p(d_i |\vt)\,.
\eeq
We notice that in this expression the only place where selection effects are present is in the term $\Nabos$. In particular, the likelihood for the data of individual sources does not include  selection effects.

Using Eq.~\refp{Eq.Ndet} one can then write
\beq
\N = \frac{\Nabos}{\alpha(\vl)}\,,
\eeq
which gives
\begin{align}
\label{Eq.LoredoLike}
p(\D | \vl \N) &= e^{-\Nabos}  \prod_{i=1}^{\Ntri} \N  \int \ud\vt\; \pi(\vt|\vl)\, p(d_i |\vt)
\notag\\ &=  e^{-\Nabos} \left(\Nabos\right)^{\Ntri}  \prod_{i=1}^{\Ntri} \frac{\int \ud\vt\; \pi(\vt|\vl)\, p(d_i |\vt) }{\alpha(\vl)}
\notag\\ &= e^{-\Nabos} \left(\Nabos\right)^{\Ntri} \alpha(\vl)^{-\Ntri}  \prod_{i=1}^{\Ntri} p(d_i | \vl)
\notag\\ &\equiv e^{-\Nabos} \left(\Nabos\right)^{\Ntri} \alpha(\vl)^{-\Ntri} \;p(D^\mathrm{tr} | \vl)\,.
\end{align}

In this expression, we have defined the set of data with triggers (which are also detections, since we are assuming no background): $D^\mathrm{tr}\equiv \{\di\}$. This must not be confused with the whose dataset $D$, which contains both segments with triggers and segments without triggers $D=(\{\di\},\{\ndj\})$, cf.~Sec.~\ref{SuSec.DataPartition}.
With the likelihood in hand, we can easily write the joint posterior of \La given the data:
\beq
p(\La | \vd) = p(\vl \N |\vd) = \frac{p(\D | \vl \N)}{p(\D|\maHL)} \pi(\N) \pi(\vl)\,,
\label{Eq.HyperPostImplicit}
\eeq
where we have used Bayes theorem and  the symbol $\pi$ for distributions that do not depend on the data. We have also written
\beq
\pi(\vl \N) = \pi(\vl|\N)\pi(\N) = \pi(\vl)\pi(\N)
\eeq
since \N and \vl are independent variables by construction in our analysis. When writing the evidence of the data, $p(\vd |\maHL)$, we have restored the explicit dependence on all settings and choices made in the hyper analysis, \maHL, since soon we will need to deal with another model.\footnote{\maHL should be on the right side of the conditional bar of all terms really, but the explicit presence of \N, \vl or \La in the other terms makes it clear which model is being used}
In this context, one often call $p(\vd |\maHL)$ the hyper evidence, a number that properly normalizes the hyper posterior:
\beq
p(\D|\maHL) \equiv \int \ud \La p(\vd | \La) \pi(\La) =\int \ud \vl \ud \N p(\vd | \vl \N) \pi(\N) \pi(\vl)\,.
\label{Eq.HyperEvidenceExpl}
\eeq
One can check that integrating Eq.~\refp{Eq.HyperPostImplicit} over \La, and using this definition of hyper evidence, returns 1, as it should be the case for a probability density.

We thus have the following key equation:
\boxit{p(\vl \N |\vd)=\frac{\pi(\N) \pi(\vl)}{p(\D|\maHL)}e^{-\Nabos} \left(\Nabos\right)^{\Ntri} \alpha(\vl)^{-\Ntri} \prod_{i=1}^{\Ntri} p(d_i | \vl)  }{HyperPosteriorFinal}
This equation depends on the number of detected triggers\footnote{Remember we already assumed that no background happens, thus \Ntri is actually a source count}  \Ntri over the experiment time and the number of sources that were expected  above the detection threshold over the experiment time \Nabos. Notice that these two numbers are not the same due to the Poissonian nature of the problem (if in doubt, see Sec.~\ref{Sec.HyperLike}!). Selection effects are contained explicitly in $\alpha(\vl)$ and implicitly in \Nabos. The individual events likelihood $p(d_i | \vl)$ will be further discussed below.
The hyper evidence, $p(\D|\maHL)$, could be dropped while doing parameter estimation, since it does not depend on the hyper parameters. It \emph{cannot} be dropped, however, if one wants to do model selection, i.e. compare how well two different models for the populations (each with its number of parameters and population priors) describe the data~\cite{LIGOScientific:2018jsj,Talbot:2018cva,Zevin:2017evb,Vitale:2015tea,Farr:2017uvj,Talbot:2017yur,Fishbach:2019bbm}. %

\subsection{Analyzing sources with different data quality}

A point worth stressing is that even though in some of the equations above (and in some of the literature) the fraction of detectable sources $\alpha(\vl)$ appears in the same product with the individual sources likelihoods, e.g., Eq.~\refp{Eq.LoredoLike}, it is \emph{not} a property of any individual source.
Instead, $\alpha(\vl)$ is a property of the population, as indicated by its dependence of the hyper parameters.

However, $\alpha(\vl)$ also depends on the sensitivity of the detector, since it is the integral over all possible source parameters produced by the population weighed by the probability that each specific value of parameters result in a detectable source, Eq.~\refp{Eq.Alpha}. But the probability that a source with given parameters \vt is detectable \emph{does} depend on the sensitivity of the detector.

What if one is including in the hierarchical analysis sources detected over a stretch of data so long that the sensitivity of the detector has changed significantly? This is for example the case for the LIGO and Virgo GW detectors in the first three science runs~\cite{2016LRR....19....1A}: sources were detected over a period of 5 years, during which the sensitivity of the detectors improved by $\sim 30$\%.

In this case one does \emph{not} calculate different $\alpha(\vl)$'s for different events. Instead, a sort of ``average''  $\alpha(\vl)$ must be used.
This can be seen from its definition Eq.~\refp{Eq.Alpha}. Let us split the parameters on which each compact binary depends, \vt, on the arrival time at Earth $t$ and all other parameters $\vec\epsilon$, then:

\beqa
\alpha(\vl) &=& \int \ud\veps \ud t \; \pi(\veps |\lambda) \pi(t| \maH)\,p(\dete|\veps\;t)\,.
\eeqa

Where we have assumed that the arrival time of the sources (and background triggers, if relevant) does not depend on the hyper parameters nor on \veps.

This can be written as
\beqa
\alpha(\vl) &=& \int \ud\veps \;\pi(\veps |\lambda)\, \overline{p}(\dete|\veps)
\,.
\eeqa
Where we have introduced the probability than a trigger with parameters \veps would be detectable averaged over the whole duration of the dataset:
\beq
\overline{p}(\dete|\veps)= \int \ud t\;\pi(t| \maH)p(\dete|\veps\;t)\,.
\eeq
If one makes the assumption that all times are equally likely\footnote{These are the times events arrive at Earth. All events, not only the detectable ones!} then one has simply:
\beq
\overline{p}(\dete|\veps)=\frac{1}{T} \int \ud t\;p(\dete|\veps\;t)\;,
\eeq
where $T$ is the total duration of the dataset.

\subsection{The hyper posterior of the shape parameters}\label{SuSec.PostShape}

Now, suppose one does not actually care about the value of the overall scale $\N$, but only about the shape parameters \vl. The posterior for those given the data can be obtained from Eq.~\refp{Eq.HyperPosteriorFinal} by marginalizing over \N:
\beqa
p(\vl | \vd) &=& \frac{\pi(\vl)}{p(\vd|\maHL)} \int \ud\N \left[e^{-\Nabos} \left(\Nabos\right)^{\Ntri} \alpha(\vl)^{-\Ntri} p(\Dt | \vl)\right] \pi(\N)\,.
\eeqa
Notice that the expected number of detectable sources \Nabos depends on the expected total number of sources \N,~Eq.~\refp{Eq.Ndet}, which we can use to write:

\beqa
p(\vl | \vd) &=& \frac{\pi(\vl)}{p(\vd|\maHL)} \int \ud\N \left[e^{-\N\alpha(\vl)} \left(\N\alpha(\vl)\right)^{\Ntri} \alpha(\vl)^{-\Ntri} p(\Dt | \vl)\right] \pi(\N) \\
&=& \frac{\pi(\vl)}{p(\vd|\maHL)}  p(\Dt |\vl) \int \ud\N \;e^{-\N\alpha(\vl)} \left(\N\right)^{\Ntri} \pi(\N)\,.
\eeqa

The marginalization can be carried out analytically in the simple case where the prior on \N is $\pi(\N) = (\N)^{-1}$:
\beqa
 \int \ud\N \;e^{-\N\alpha(\vl)} \N^{\Ntri} \pi(\N) &=&  \int \ud\N \;e^{-\N\alpha(\vl)} \frac{\left(\N\right)^{\Ntri} }{\N} \\ =
\int \ud\N \;e^{-\N\alpha(\vl)} \N^{\Ntri-1}  &\overset{\mathrm{y\equiv \N \alpha}}{=}& \int \frac{\ud y}{\alpha(\vl)}\;e^{-y} \left[\frac{y}{\alpha(\vl)}\right]^{\Ntri-1}\\
= \alpha(\vl)^{-\Ntri}  \int {\ud y} \;e^{-y} y^{\Ntri-1} &=& \alpha(\vl)^{-\Ntri} \;\Gamma(\Ntri -1)\,,
\eeqa
where $\Gamma$ is the gamma function. Plugging this back we have:

\beq
p(\vl | \vd) =  \frac{\pi(\vl)}{p(\vd|\maHL)}  p(\Dt|\vl)  \alpha(\vl)^{-\Ntri} \;\Gamma(\Ntri -1)\,.
\eeq
Remembering that $p(\Dt |\vl)$ is the product the \Ntri single-event likelihood we have:
\beq
p(\vl | \vd)=  \frac{\pi(\vl)}{p(\vd|\maHL)}  \;\Gamma(\Ntri -1)  \prod_{i=1}^{\Ntri} \frac{p(d_i |\vl) }{\alpha(\vl)}\,. \label{Eq.HyperPostOfLikeFirst}
\eeq
To calculate this expression, we need for each source the likelihood of the data given \vl, $p(d_i |\vl)$. This can be written in a more useful form by using the sum rule to marginalize over all possible values of the source parameters:

\beqa
p(\vl | \vd)&=& \frac{\pi(\vl)}{p(\vd|\maHL)}  \;\Gamma(\Ntri -1)  \prod_{i=1}^{\Ntri} \frac{\int \ud \vt p(d_i |\vt\, \vl) \pi(\vt|\vl)}{\alpha(\vl)} \\
 &=& \frac{\pi(\vl)}{p(\vd|\maHL)}  \;\Gamma(\Ntri -1)  \prod_{i=1}^{\Ntri} \frac{\int \ud \vt p(d_i |\vt ) \pi(\vt|\vl)}{\alpha(\vl)}\,.
 \label{Eq.HyperPostOfLike}
\eeqa
In the last step we have eliminated \vl from the right side of the conditional bar of $p(d_i |\vt \vl)$ since the likelihood only depends on the source parameters, and not on the population hyper parameters.
Also notice  how the selection effects are now entirely contained in $\alpha(\vl)$ (cf. Eq.~(20) of ~\cite{2018arXiv180510270F}).

How do we calculate the likelihood for the i-th data segment, $p(d_i |\vt )$?
The most common scenario is that we have a set of posterior samples $p(\vt |d_i )$, produced by either ourselves, a colleague, or a scientific collaboration, using  a Markov Chain Monte Carlo (MCMC) or similar algorithm, since in general these are high-dimensional non-trivial distributions to sample~\cite{2015PhRvD..91d2003V,Ashton:2018jfp}.
If the sampling prior used to generate those samples is $\pi(\vt| \maHP)$ then posterior and likelihood are related as:
\boxit{p(d|\vt) =p(\vt|d,\maHP)\frac{p(d|\maHP)}{\pi(\vt |\maHP)}\,,}{Bayes}
where we have used \maHP (``PE'' stands for Parameter Estimation) for the modeling used to characterize the individual events and  $p(d|\maHP)$ is the evidence of the data obtained with the sampling algorithm:
\beq
p(d | \maHP) \equiv \int \ud \vt \; p(d| \vt) \pi(\vt| \maHP)\,.
\label{Eq.SingleLike1}
\eeq

Notice that \maHP and \maHL imply different prior beliefs in the distribution of \vt:
the former model assumes that the parameters \vt are distributed as $\pi(\vt|\maHP)$, i.e. the sampling prior, while the latter assumes  $p(\vt| \vl)$, i.e. the population prior.
This means that  in general the hyper evidence for the triggers is not equal to the evidence calculated by the sampling algorithm:
\beq
\prod_{i=1}^{\Ntri} p(d_i | \maHL) \neq \prod_{i=1}^{\Ntri} p(d_i | \maHP)\;.
\label{Eq.UnequalEvidences}
\eeq

A last point worth stressing before we continue is the following. \textit{If} the sampling prior $\pi(\vt | \maHP)$ is uniform in \textit{all} parameters then:
\beq
\pi(\vt|\maHP)= c \implies p(d|\vt) \propto p(\vt|d)\,,
\label{Eq.FlatPriorMagic}
\eeq
and one could use the draws from the posterior as if they were draws from the likelihood function. Unfortunately, in general, at least some of the sampling priors are non-trivial, and must be taken into account when going from posterior to likelihood, as we saw in Eq.~\refp{Eq.Bayes}  (one often colloquially says that they need to ``remove'' the prior when referring to the reweighing required by Eq.~\refp{Eq.Bayes}).

With this in mind, let's use Bayes theorem inside the product of Eq.~\refp{Eq.HyperPostOfLike}:
\beqa
p(\vl | \vd) &=&\frac{\pi(\vl)}{p(\vd|\maHL)}  \;\Gamma(\Ntri -1)  \prod_{i=1}^{\Ntri} \frac{\int \ud \vt p(d_i |\vt ) \pi(\vt|\vl)}{\alpha(\vl)}\notag \\
 &=& \frac{\pi(\vl)}{p(\vd|\maHL)}  \;\Gamma(\Ntri -1)  \prod_{i=1}^{\Ntri}\frac{1}{{\alpha(\vl)}} \int\ud \vt \;\;\frac{ p(\vt | d_i) p(d_i|\maHP) \pi(\vt|\vl) }{\pi(\vt|\maHP)}\notag\\
 &=&\pi(\vl) \;\Gamma(\Ntri -1) \frac{p(\Dt| \maHP)}{p(\D|\maHL)} \prod_{i=1}^{\Ntri} \frac{1}{{\alpha(\vl)}} \int \ud \vt\;\; \frac{  p(\vt | d_i)\pi(\vt|\vl) }{\pi(\vt|\maHP)}\label{Eq.PostLambdaOfPost}
\eeqa
where in the first equation we have used the fact that the single event likelihood does not  depend on \vl. In the second row $\pi(\vt|\maHP)$ is the sampling prior for \vt used when drawing posterior samples (using e.g. an MCMC sampler, as mentioned above). The ratio of the MCMC evidence over the hyper evidence is ugly to see, but luckily is not really important as long as we are only trying to measure the population hyper parameters of \emph{one} model \maHL and not comparing different models, since this ratio does not depend on the actual value of the hyper parameters. This is why it is only seldom explicitly reported in the literature.

The integral in Eq.~\refp{Eq.PostLambdaOfPost} can be simplified by noticing that it represents the expectation of the ratio of the population prior over the sampling prior.  The exact expectation can be replaced with the approximate sample mean~\cite{mackay2003information}, i.e. with a discrete sum:
\beq
\int \ud \vt \; p(\vt | \maH) f(\vt) \simeq \frac{1}{\Nsa} \sum_{j=0}^{\Nsa} \left. f(\vt^j) \right|_{\vt^j\sim p(\vt| \maH)}\,,
\eeq
which is valid for any function $f(\vt)$ (the symbol $\sim$ in this context means ``drawn from the distribution''), as long as the number of available posterior samples \Nsa is large enough (the larger \Nsa, the better the approximation).

Therefore, assuming we are given $\Nsa$ posterior samples from the posterior distribution of \vt for each of the \Ntri sources, then we can write (cfr Eq. (14) of ~\cite{Mandel:2018mve}):
\beq
\prod_{i=1}^{\Ntri}\frac{1}{\alpha(\vl)}\!\! \int \! \ud \vt p(\vt | d_i)\frac{\pi(\vt|\vl) }{\pi(\vt|\maHP)} = \prod_{i=1}^{\Ntri} \frac{1}{\Nsa\; \alpha(\vl)} \!\!\!\sum_{j=0}^{\Nsa}\!\!\!\!\! \left. \frac{\pi(\vt^j_i|\vl) }{\pi(\vt^j_i|\maHP)}\right|_{\vt^j_i \sim p(\vt | d_i ,\maHP)},
\label{Eq.Mandel14}
\eeq
where $\vt^j_i$ is the j-th posterior sample from the posterior distribution of the i-th source.

Since we are calculating an average, one often uses angle brackets:
\beq
 \frac{1}{\Nsa} \sum_{j=0}^{\Nsa} \left. \frac{p(\vt^j_i|\vl) }{\pi(\vt^j_i|\maHP)}\right|_{\vt^j_i \sim p(\vt | d_i ,\maHP)}=\left<\frac{\pi(\vt_i|\vl) }{\pi(\vt_i|\maHP)} \right>_{\vt_i \sim p(\vt | d_i ,\maHP)}\,.
\eeq

So finally we obtain:
\boxit{p(\vl | \vd) =\pi(\vl) \;\Gamma(\Ntri -1)\; \alpha(\vl)^{-\Ntri} \; \frac{p(\Dt| \maHP)}{p(\D|\maHL)} \prod_{i=1}^{\Ntri}\left<\frac{\pi(\vt_i|\vl) }{\pi(\vt_i|\maHP)} \right>_{\vt_i \sim p(\vt | d_i,\maHP) }.}{HyperPost}
It is worth stressing and remembering that the selection effects are here included in the detectable fraction $\alpha(\vl)$, and that this term comes from marginalizing over the total number of sources, not from the individual events likelihood.

\subsection{Handling missing parameters}\label{SuSec.WhatWeDontSay}

Suppose that the detected signals can be characterized by two unknown parameters $\vt=(x,y)$, but we really only care about $x$ because it is the parameter that will tell us about the underlying population parameters \vl. In other words, the distribution of the $x$ parameters of the population of events does depend on $\vl$, $\pi(x | \vl ,\maHl)$, but we believe the distribution of $y$ is fixed and known: $\pi(y|\maHl)$. For example, we might be confident that the sky positions of the GW sources we detected is isotropic, and do not wish to include them in the hierarchical analysis.
Or, we might be using someone else's public data release, but they only gave us the marginalized $x$ posteriors, not the whole cloud of points.

Can we calculate the \vl posterior by only integrating over $x$, and neglecting $y$?
Looking at Eq.~\refp{Eq.HyperPostOfLikeFirst} one might be tempted to use the sum rule and write:
\beqa
p(d_i | \vl) &=& \int \ud x\; p(d_i x | \vl) = \int \ud x\; p(d_i | x)\pi(x | \vl) \nonumber\\
&=& \int \ud x\; p( x|d_i ) p(d_i) \frac{\pi(x | \vl,\maHl)}{\pi(x|\maHP)} \nonumber\\
&=& p(d_i) \left\langle \frac{\pi(x | \vl,\maHl)}{\pi(x|\maHP) }\right\rangle_{x\sim p(x|d_i)}\,,\label{Eq.WrongOneDLike}
\eeqa
where $\pi(x|\maHP)$ is the sampling prior on $x$ used when sampling the posterior distribution. Now, this seems a perfectly legitimate equation; but one could also integrate on both x and y and obtain:
\beq
p(d_i | \vl) = \int \ud x\,\ud y\; p(d_i\, x\, y | \vl) = p(d_i) \left\langle \frac{\pi(x | \vl,\maHl) \pi(y|\maHl)}{\pi(x y |\maHP) }\right\rangle_{x,\,y\sim p(x, y|d_i)}\label{Eq.RightTwoDLikeUnexpanded}
\eeq
Let's assume for simplicity that the sampling prior factorizes, $\pi(x y |\maHP)=\pi(x |\maHP) \pi(y |\maHP)$:
\beq
p(d_i | \vl) = p(d_i) \left\langle \frac{\pi(x | \vl,\maHl) \pi(y|\maHl)}{\pi(x |\maHP) \pi(y |\maHP) }\right\rangle_{x,\,y\sim p(x, y|d_i)}\label{Eq.RightTwoDLike}
\eeq

Clearly, the two expressions in the last term of Eq.~\refp{Eq.WrongOneDLike} and Eq.~\refp{Eq.RightTwoDLike} are in general different, because the values of $x$ used in the two cases are the same, no matter of whether we use the samples from the 2D $x\sim p(x\,y| d_i)$ or from the marginalized distribution $x\sim p(x|d_i)$, but now there is an extra term $\pi(y|\maHl)/\pi(y|\maHP)$ which need not be equal to 1.
On the other hand, the left hand sides look the same. What is happening?  The problem is that we have not explicitly written down \emph{all} of the assumptions on the right side of the conditional bars.

What we want for the hierarchical analysis, Eq.~\refp{Eq.HyperPostOfLikeFirst}, is really the probability of the data given $\vl$ \emph{and} our model \maHl. The model comes with set of priors on the distribution of  population parameters, \textit{all} parameters:
\beq
p(d_i | \vl) =p(d_i | \vl \maH_{\vl}) = p(d_i | \vl \land \pi(\vt|\lambda))= p(d_i | \vl \land \pi(x|\vl,\maHl) \land \pi(y|\maHl))\,.
\label{Eq.ExplicitTwoD}
\eeq
In fact, this is the very reason why we have to ``divide out'' the sampling priors in Eq.~\refp{Eq.RightTwoDLike}: we want to calculate this likelihood under the hypothesis that the prior being used is the population one $\pi(\vt | \vl,\maHl)$, but we have been handled samples that were produced using the prior $\pi(\vt|\maHP)$.
Instead, in Eq.\refp{Eq.WrongOneDLike} we calculated:
\beq
p(d_i | \vl)  = p(d_i | \vl \land \pi(x|\lambda \maHl))\,.
\eeq

Because nothing is said about the prior on the other variable, this is implicitly assuming that the prior on $y$ is \emph{still the one used while sampling} the parameter space, i.e. $\pi(y| \maHP)$:
\beq
\mathrm{Eq.\refp{Eq.WrongOneDLike}} \longleftrightarrow p(d_i | \vl \land \pi(x|\lambda, \maHl) \land \pi(y|\maHP))
\label{Eq.ExplicitOneD}
\eeq
This is why the right hand sides of Eq.\refp{Eq.WrongOneDLike} and Eq.~\refp{Eq.RightTwoDLike} look like they are going to give two different results: in general, they will!

The only parameters that we could not explicitly account for while calculating the posterior of the  population hyper parameters are those for which the population prior is \emph{exactly equal} to the sampling prior. This can be seen directly comparing the two equations above: the LHSs [Eq.~\refp{Eq.ExplicitOneD} and Eq.~(\ref{Eq.ExplicitTwoD})] and the RHSs [Eq.\refp{Eq.WrongOneDLike} and Eq.~(\ref{Eq.RightTwoDLike})] become identical if $\pi(y|\maHl)=\pi(y|\maHP)$.\footnote{Notice that the evidences on the right hand side of Eq.\refp{Eq.WrongOneDLike} and Eq.~\refp{Eq.RightTwoDLike} are also in general different, for the same reason.}
But notice that this still implies that the relevant prior distributions are separable, $\pi(x y |\maHP)=\pi(x |\maHP) \pi(y |\maHP)$, which might not always be the case! If the priors are more complicated, and correlations exist, it is even hard to simplify things so that only some of the prior distributions survive.
Therefore Eq.~\refp{Eq.RightTwoDLikeUnexpanded} is \emph{always} what we want to do, while Eq.~\refp{Eq.WrongOneDLike} \emph{might} be sometimes what we want to do, if we are very careful and lucky.

One might still wonder: why should we be careful if a constant (i.e. not depending on \vl, which is the quantity we are interested in) multiplicative term is missing? Can't this be accounted for later, while enforcing the proper normalization of the \vl posterior?
The answer is no, this would create an issue. This can be seen best by looking at the RHS of ~Eq.~\refp{Eq.Mandel14}, which in our 2D example reads:
\beq
\prod_{i=1}^{\Ntri} \frac{1}{\Nsa} \sum_{j=0}^{\Nsa} \left. \frac{\pi_i(x^j |\vl  \maHl) \pi(y^j|\maHl)}{\pi(x^j|\maHP) \pi(y^j |\maHP)} \right|_{x^j,\,y^j\sim p(x, y|d_i)}\,
\eeq
where we are neglecting the $\alpha$ term, which is not important for this discussion, and assuming all priors factorize.

If $\pi(y^j|\maHl) \neq \pi(y^j |\maHP)$ in general the ratio $\pi(y^j|\maHl)/ \pi(y^j |\maHP)$ will be different from 1, and it will be different for different samples and different events. Thus if we entirely neglected $y$, we would be adding a different, and wrong, numerical weighting coefficient for each sample in these sums.

What if someone is only handled marginalized posteriors? For example, the LIGO~\cite{Harry:2010zz}-Virgo~\cite{TheVirgo:2014hva} collaboration (LVC) currently only releases in low latency a estimate of the sky position and distance, but not of the masses, of the candidate compact binary coalescence (CBC) sources it detects~\cite{LVCUser}. In absence of this information one cannot correctly remove the mass prior, hence their inference on any hyper parameters, even if not directly linked to the masses, would be biased unless their population prior for the masses were identical to what the LVC used to produce samples (and assuming prior factorizes). If that is the case, then masses can be entirely dropped out while calculating the likelihood term above, but, again,  care must be taken to be sure that the very same prior is used as the population prior when calculating selection effects, i.e. $\alpha(\vl)$, or else biases will be introduced.
A common mistake is to use posterior samples which have already been marginalized over some parameter that is not of interest (e.g. $y$ in the example above), hence implicitly using a prior for $y$ equal to the sampling prior, but then use a \emph{different} prior when calculating $\alpha(\vl)$.

\section{A simple 1-dimensional example}\label{Sec.OneDExample}

We know consider a simple but instructive 1D example, that shows how many of these pieces come to play, in practice.

\begin{figure}[t]
\centering
\if\blackbgd1
\includegraphics[width=0.75\textwidth]{Speakers_bw.png}
\else
\includegraphics[width=0.75\textwidth]{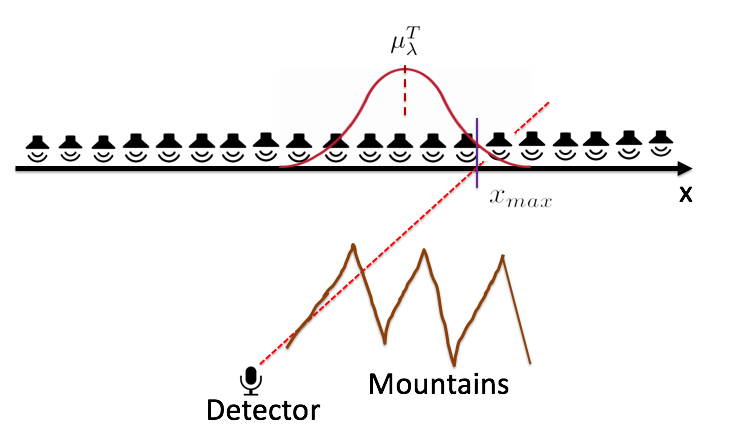}
\fi
\caption{A sketch of the problem described in Sec.~\ref{Sec.OneDExample}. The red curve represents the density of speakers that are playing. %
}\label{Fig.Speakers}
\end{figure}

The problem is as follows. There is an infinite number of audio speakers arranged along the x-axis. An audio detector is placed at some distance as in Fig.~\ref{Fig.Speakers}. When one speaker at position $x$ emits a sound, %
the detector records:
\beq
d= x + n
\eeq
where $x$ is the true position of the speaker that emitted a signal, and $n \sim \mathcal{N}(0,1)$ is a random number drawn from a normal distribution  with mean equal to 0 and standard deviation equal to 1 meter. Here we are just assuming, arbitrarily, that the noise associated with recording the data has this typical width of 1 meter. We can think of this noise term as due to e.g. winds that affect the perceived direction of the incoming audio wave.

More realistic cases will differ from this in a few ways:
\begin{enumerate}
\item the noise properties (e.g. its standard deviation) might be time dependent;
\item noise might be different for different events
\item data might not directly contain the variable of interest, but rather a function of it (i.e. $d= n + h(x)$, with $h$ some potentially complicated function, see e.g.  Sec.~\ref{Sec.GWExample}); and
\item there might be more than one parameter that affects the detected data, all or some of which might be of interest to us, and all or some of which might depend on a higher hyper-level.
\end{enumerate}

In this example, because of a nearby mountain range, there exists a maximum value of $x$ such that detectors placed at $x>x_{\rm max}$ would not produce a measurable sound.
However, it is important to remember that \textit{we do not know x!}. The best we can do is to use the data to set a threshold. Since the noise in our example is a zero-mean process it's reasonable to use as a detection criteria:
\beq
 d \leq x_{\rm max} \longleftrightarrow \rho(d)>\rhot\,.
 \label{Eq.DetCriterium}
\eeq

But we should not forget that there is noise! Thus even a source with $x_{\rm true}=x_{\rm max}$ will produce a detectable signal $50\%$ of the time (50\% since the noise is symmetric).
Let's pause for a second and stress a couple of important points:
\begin{itemize}
\item Barring rare exceptions, the selection criteria is based on the data, not on the true value of the parameter. In general we do not know the parameters! The data is all we have. The data are our friend.
\item Because of the noise, sources with a \emph{true} value of the parameter that fail the threshold for detection can be promoted to detections. Conversely, sources that would meet the criteria, can fail it because of the noise.
\end{itemize}
This also means that unless there is absolutely no noise (i.e. the noise is so small that we are de facto measuring the true value of the parameters), the selection effect will not be  sharp feature, a Heaviside function in our simple 1D case, but rather a sigmoid that smoothly turns off.

Let's see this explicitly. First, given that the noise is gaussian, the likelihood for the data $d$ given $x$ is:

\beq
p(d|x) =\frac{1}{\sqrt{2 \pi}} \exp{\left[ -\frac{1}{2} (d-x)^2\right]}\,
\label{Eq.SingleEvLike}
\eeq
which contains, implicitly, also the standard deviation of the noise, which is equal to 1 meter. Thus in our scenario the likelihood for the event is a gaussian, seen as a function of $x$, centered at $d$.

We had previously defined the probability of a source being detectable given the true value of the parameter, Eq.~\refp{Eq.PDet}:
\beq
p(\dete | x) =  \int_{\md_\uparrow} \ud d\;\; p(d|x)\,.
\label{Eq.PDetX}
\eeq
This integral is made over the detectable data, which are the data for which $d<\xmax$, i.e.:
\begin{align}
p(\dete | x) &=  \int_{\md_\uparrow} \ud d\; p(d|x) \notag \\&= \int_{-\infty}^{\xmax} \ud d\; \frac{1}{\sqrt{2 \pi}} \exp{\left[ -\frac{1}{2} (d-x)^2\right]} = \frac{1}{2}\left[1+ \mathrm{erf}\left(\frac{\xmax-x}{\sqrt{2}}\right)\right]\,.
\end{align}

\begin{figure}
\centering
\if\blackbgd1
\includegraphics[width=0.5\textwidth]{pdet_bw.pdf}
\else
\includegraphics[width=0.5\textwidth]{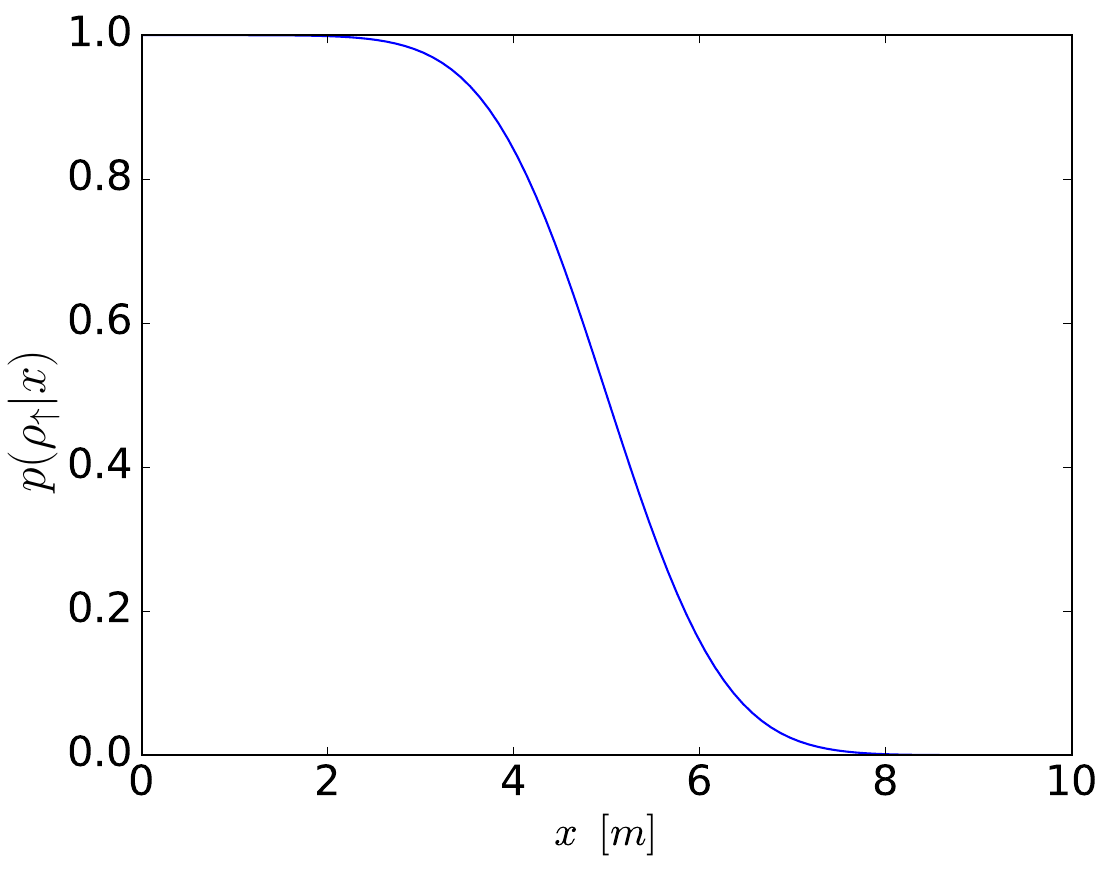}
\fi
\caption{The probability $p(\dete |x)$ of detecting the sound from a speaker at true position $x$. Here $\xmax=5$}\label{Fig.Pdet}
\end{figure}

We see that $p(\dete | \xmax) =0.5$, as mentioned above: half of the sources with true value of $x$ at threshold are detectable because of the noise. The overall curve is indeed a sigmoid, as reported in Fig.~\ref{Fig.Pdet}.

Let us pause here for a second to stress an important point: we said earlier that the \emph{criteria} we use to decide if a piece of data contains a trigger must only depend on the data, not on true value of the parameter, Eq.~\refp{Eq.DetCriterium}. However, the probability that any given value of $x$ results in a detection \emph{does} depend on $x$, Eq.~\refp{Eq.PDetX}.
There is no contradiction: in Eq.~\refp{Eq.PDetX} $x$ is on the right hand side of the conditional bar, and hence it's assumed known. Whether and when that particular $x$ result in a detection is still determined by the data: given $x$ one can calculate all possible data that can be contain it, i.e. all possible noise realizations, and adds up those that result in a detectable piece of data.

Let us now consider inference on the property of the speakers, given a set of detections. We will assume that a friend of ours has setup the system such that over a long period of time\footnote{Long enough that the probability that two speakers play in the same small data chunk is negligible and we can use the expressions we developed so far.} a set of speakers will play, one at the time. The positions of the speakers that will play are scattered around some central value $\mut$, and are draws from a normal distribution (cf. Fig.~\ref{Fig.Speakers}):
\beq
\pi(x | \mu_\lambda, \sigma_\lambda) = \mathcal{N}(x-\mu_\lambda, \sigma_\lambda).
\eeq
Our friend has told us the value of $\sigma_\lambda$, which we can thus treat as known, but has challenged us to measure the true value of \mul, \mut, given the data recorded by the detector.

We have all the tools we need to meet the challenge. The mathematics has been laid out in the previous sections, and can be directly reused with the understanding that our only shape  hyper-parameter\footnote{We are not trying to infer the total number of detectors that played.} is:

$$\vl =\{\mul\}$$

\begin{figure}
\centering
\if\blackbgd1
\includegraphics[width=0.5\textwidth]{alpha_bw.pdf}
\else
\includegraphics[width=0.5\textwidth]{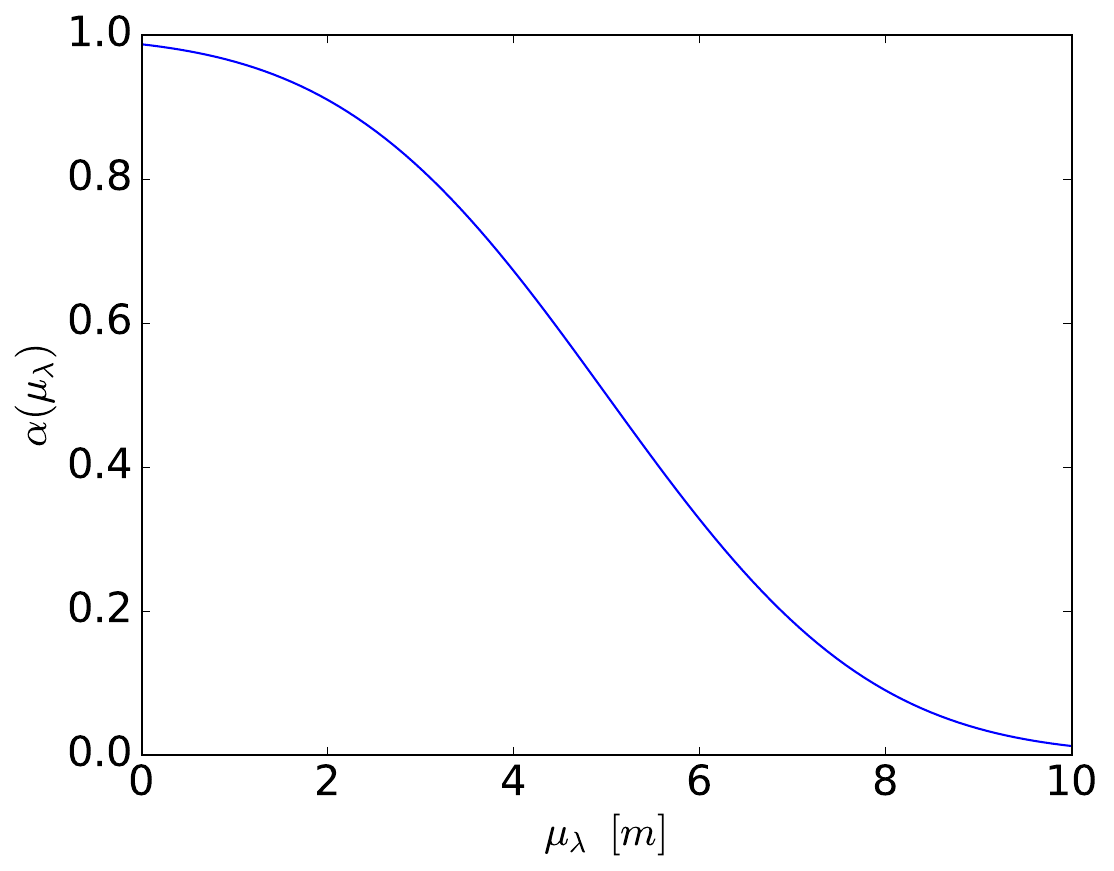}
\fi
\caption{The fraction of detectable events $\alpha(\mul)$ as a function of the hyper parameter \mul. Here $\sil =2$ and $\xmax=5$}\label{Fig.Alpha}
\end{figure}

The posterior on \mul is given by Eq.~\refp{Eq.HyperPostOfLike}. We first need to calculate the detection efficiency $\alpha(\vl)$:
\beq
\alpha(\mul)=  \int \ud x\; \pi(x |\mul) p(\dete|x)
\eeq
which is shown in Fig.~\ref{Fig.Alpha} for $\mut=4.3$, $\sil=2$ and $\xmax=5$.

With this in hand, we consider the pieces of  Eq.~\refp{Eq.HyperPostOfLike} that depend on $\mul$, and neglect everything else since it will only act as a constant term with respect to \mul (which we can be fixed by normalizing the posterior at the end). We have:
\beq
p(\mul | \vd)\propto   \pi(\mul)  \prod_{i=1}^{\Ntri} \frac{\int \ud x \; p(d_i |x ) \pi(x|\mul)}{\alpha(\mul)}\,.
\label{Eq.MulPostLike}
\eeq
Because both the individual event likelihood and the population prior are Gaussians, we can actually solve the numerator analytically %
\beqa
\int \ud x \; p(d_i |x ) \pi(x|\mul) &=&  \frac{1}{2 \pi \sil} \int \ud x \;  e^{\left[ -\frac{1}{2} (d_i-x)^2\right]}e^{\left[ -\frac{1}{2\sil^2} (x-\mul)^2\right]}\,.
\eeqa

Before we continue, let us stress again that in here the integration range is the \emph{full and unadulterated} range that $x$ can theoretically take, %
{\emph{including}} values which are above our threshold \xmax. As seen before, the selection effects do not enter the numerator of the hyper posterior, but only though $\alpha(\vl)$.

In this case, this is of great help because it allows us to get rid of the integral using standard Gaussian integrals. In fact if
\beq
f(x)=\frac{1}{\sqrt{2 \pi} \sigma_{f}} e^{-\frac{\left(x-\mu_{f}\right)^{2}}{2 \sigma_{f}^{2}}}
\qquad {\rm and}\qquad
g(x)=\frac{1}{\sqrt{2 \pi} \sigma_{g}} e^{-\frac{\left(x-\mu_{g}\right)^{2}}{2 \sigma_{g}^{2}}}\,,
\eeq
then
\beq
f(x) g(x)=\frac{A}{\sqrt{2 \pi} \sigma_{f g}} \exp \left[-\frac{\left(x-\mu_{f g}\right)^{2}}{2 \sigma_{f g}^{2}}\right]
\eeq
where
\beq
\mu_{f g}\equiv \frac{\mu_{f} \sigma_{g}^{2}+\mu_{g} \sigma_{f}^{2}}{\sigma_{f}^{2}+\sigma_{g}^{2}}\,,
\qquad
\sigma_{f g}\equiv \sqrt{\frac{\sigma_{f}^{2} \sigma_{g}^{2}}{\sigma_{f}^{2}+\sigma_{g}^{2}}}\,,
\eeq
and
\beq
A\equiv \frac{1}{\sqrt{2 \pi\left(\sigma_{f}^{2}+\sigma_{g}^{2}\right)}} \exp \left[-\frac{\left(\mu_{f}-\mu_{g}\right)^{2}}{2\left(\sigma_{f}^{2}+\sigma_{g}^{2}\right)}\right]
\nonumber\,.
\eeq

In our case this gives:
\beq
\frac{1}{2 \pi \sil} \int \ud x \;  e^{\left[ -\frac{1}{2} (d_i-x)^2\right]}e^{\left[ -\frac{1}{2\sil^2} (x-\mul)^2\right]} = \frac{1}{\sqrt{2 \pi\left(1+\sil^{2}\right)}} \exp \left[-\frac{\left(d_i-\mul \right)^{2}}{2\left(1+\sil^{2}\right)}\right]
\eeq
where we have used the standard Gaussian integral
\beq
\frac{1}{\sqrt{2 \pi} \sigma_{f g}} \int \ud x\;\exp \left[-\frac{\left(x-\mu_{f g}\right)^{2}}{2 \sigma_{f g}^{2}}\right]=1
\eeq

We thus see that each detection contributes to the overall posterior for \mul with a term proportional to:
\beq
\exp \left[-\frac{\left(d_i-\mul \right)^{2}}{2\left(1+\sil^{2}\right)}\right]
\label{Eq.SingleLike}
\eeq
and the overall posterior for \mul is:
\boxit{
p(\mul | \vd)\propto   \pi(\mul)  \prod_{i=1}^{\Ntri} \frac{\exp \left[-\frac{\left(d_i-\mul \right)^{2}}{2\left(1+\sil^{2}\right)}\right]}{\alpha(\mul)}
}{MulFinal}

The width of the \mul posterior is clearly determined by how many detections we have, with more events resulting in a sharper peak.
Similarly, smaller \sil help, since a) the smaller \sil the more the typical data will be close to \mut, making more events contribute useful information b) the denominator in the single-event exponent will be smaller, meaning that the tails of the single-event likelihoods will contribute less and less. If we had not fixed the noise width of each event to be the same and equal to 1 meter, a smaller width would also make the inference better, for the same identical reasons.

We notice that the term $1/\alpha(\mul)$ gives a boost to \mul's which result in more hard-to-detect events. Of course, the price to pay for making the denominator smaller is that the numerator will also decrease if we are considering \mul's that are far from our data.
The \mul posterior is thus determined by the delicate balance between these two factors.

 In this simple example we have actually integrated the likelihood analytically, to obtain Eq.~\refp{Eq.MulFinal}.
 A possible implementation of this algorithm would thus be:

 \begin{itemize}
 \item Choose the values of \mut, \sil, \xmax which will be used to generate the synthetic set of measurements.
 \item Generate $N$ random numbers  $\vec{x}_{T} \sim \mathcal{N}(\mut,\sil)$, those are the true positions of the speakers that will be playing, and we do not have access to those directly.
\item Generate $N$ noisy data sets by adding to each $x$ in $\vec{x}_{T}$ a random draw from a normal distribution: $\vec{\xi} =\mathcal{N}(\vec{x}_{T} ,1m)$ with unit standard deviation.
\item Only retain the  data that satisfy the detection criteria:
\beq
\forall d_i \in \vec{\xi},  d_i \in D \quad \mathrm{  if  }  \quad d_i <\xmax
\eeq
where
$ D$ is  our catalog of \Ntri detected sources that will be used for the inference.

\item Grid the \mul axis with enough points, and at each point calculate $\alpha(\mul)$ and Eq.~\refp{Eq.SingleLike} for each event.

\item Pick a prior for \mul, flat is just fine, and calculate Eq.~\refp{Eq.MulFinal}. Note that, because we marginalized over $x$ analytically, we do not actually have to create synthetic posteriors for the individual events - only their measured data $d_i$ matters, see Eq.\refp{Eq.MulFinal}.
 \end{itemize}

The reason why we managed to proceed so far using analytical calculations is because the problem in hand is quite simple.  In general the problem will be too complicated for the maths to be carried out analytically, due to higher dimensionality, correlations, non-trivial priors, etc. In that case, a good starting point is Eq.~\refp{Eq.MulPostLike}:
\beqa
p(\mul | \vd)&\propto&   \pi(\mul)  \prod_{i=1}^{\Ntri} \frac{\int \ud x \; p(d_i |x ) \pi(x|\mul)}{\alpha(\mul)} 
\notag\\
&\propto&   \pi(\mul)  \prod_{i=1}^{\Ntri} \frac{\int \ud x \; p(x|d_i ) \frac{\pi(x|\mul)}{\pi(x|\maHP)}}{\alpha(\mul)} \notag\\
&\propto&   \pi(\mul)  \prod_{i=1}^{\Ntri} \frac{1}{\alpha(\mul)} \left\langle \frac{\pi(x_i | \mul)}{\pi(x_i|\maHP)}\right\rangle_{x_i\sim p(x|d_{i})}\,,
\eeqa
where in the last equality we are averaging the ratio of the population prior over the sampling prior [i.e. the prior used to produce the posterior samples $p(x | d_i)$] calculated at values of $x$ fairly drawn from the posterior samples.

Assuming that for each source a set of $\Nsa$ samples have been drawn from the posterior of each trigger then we can write this as:
\beq
p(\mul | \vd) \propto \frac{ \pi(\mul)}{\alpha(\mul)^{\Ntri}}  \prod_{i=1}^{\Ntri} \frac{1}{\Nsa}  \sum_{j=1}^{\Nsa} \left. \frac{ \pi(x^j_i | \mul) }{ \pi(x^j_i|\maHP)} \right|_{x^j_i\sim p(x|d_{i})}\,.\label{Eq.MulPostMean}
\eeq

If we want to do this with our problem, and generate samples instead of using the analytic integration over $x$, we would do (see Appendix A of ~\cite{Fishbach:2019ckx} for a more complicated application):

 \begin{itemize}
 \item Choose the values of \mut, \sil, \xmax which will be used to generate the synthetic set of measurements.
 \item Generate $N$ random numbers  $\vec{x}_{T} \sim \mathcal{N}(\mut,\sil)$, those are the true positions of the speakers that will be playing, and we do not have access to those directly.
\item Generate $N$ noisy data sets by adding to each $x$ in $\vec{x}_{T}$ a random drawing from a normal distribution: $\vec{\xi} =\mathcal{N}(\vec{x}_{T} ,1m)$
\item Only retain the data that satisfy the detection criteria:
\beq
\forall \; d_i \in \vec{\xi},  d_i \in D\quad \mathrm{  if  }  \quad d_i <\xmax
\eeq
where $D$ is  our catalog of \Ntri detected sources that will be used for the inference.

\item For each $d_i \in D$ draw \Nsa samples from the likelihood distribution Eq.~\refp{Eq.SingleEvLike}:
\beq
\{d_i^1, d_i^2\ldots {d_i}^{\Nsa}\} \sim \mathcal{N}(d_i; 1\mathrm{m})\,.
\eeq
\item If a flat sampling prior on $x$ is being assumed, then these are also samples from the posterior of $x$, and can ben used directly in Eq.~\refp{Eq.MulPostMean}. Otherwise importance resampling, or an analogous technique, must be used.

\item Pick a prior for \mul, flat is just fine, grid the \mul axis with enough points, and at each point calculate $\alpha(\mul)$ and then Eq.~\refp{Eq.MulPostMean}
\end{itemize}

\begin{figure}[t]\centering
\if\blackbgd1
\includegraphics[width=0.7\textwidth]{Speakers_43_bw.pdf}
\\\vspace{0.2cm}
\includegraphics[width=0.7\textwidth]{Speakers_83_bw.pdf}
\else
\includegraphics[width=0.7\textwidth]{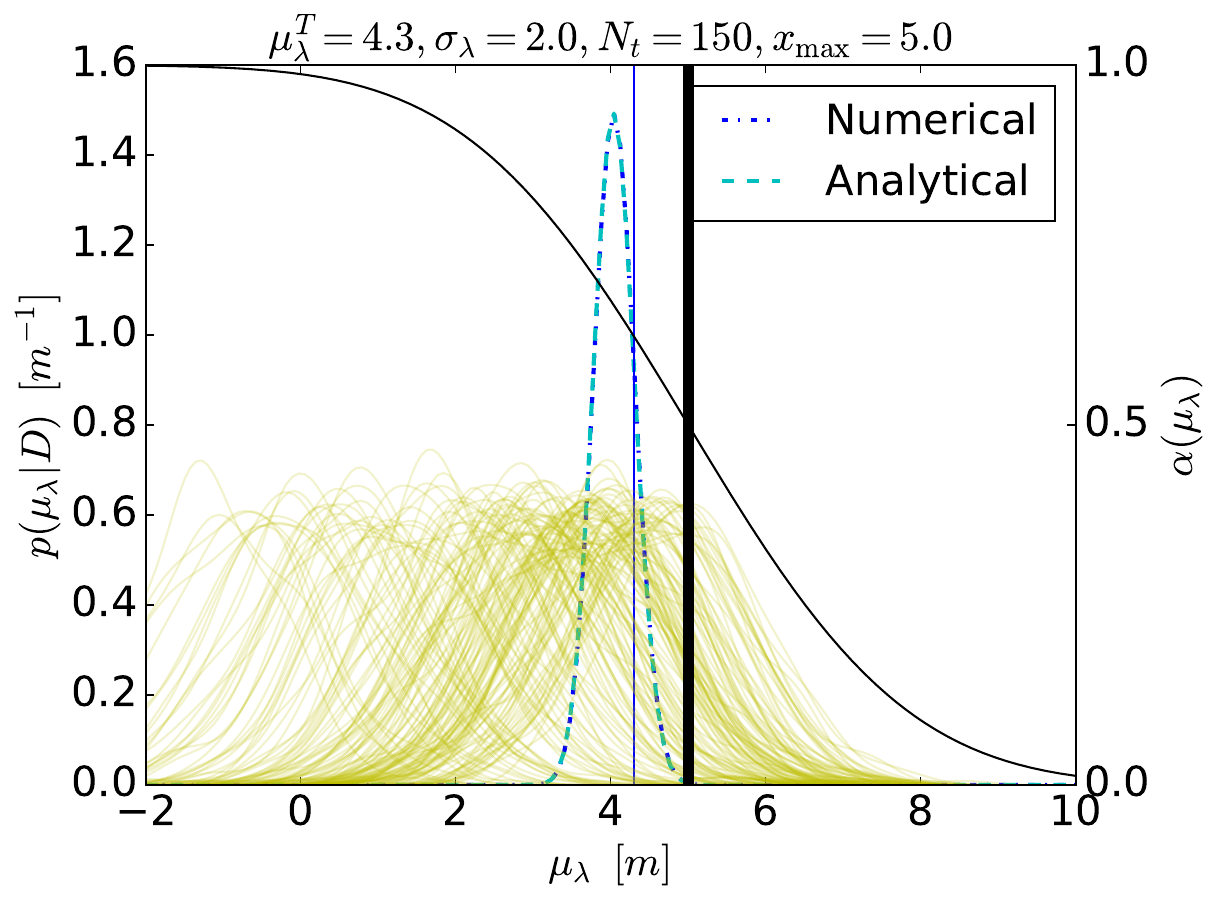}
\\\vspace{0.2cm}\includegraphics[width=0.7\textwidth]{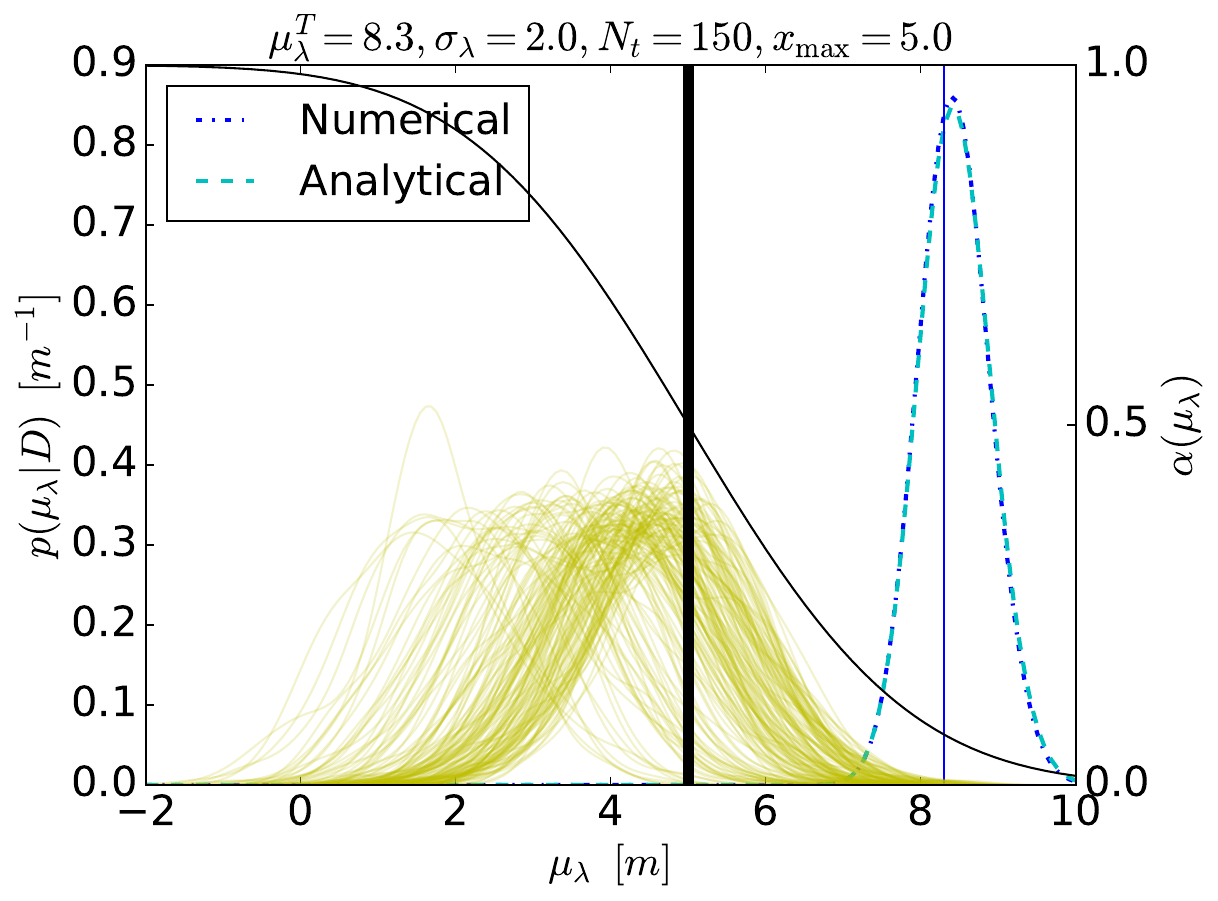}
\fi
\caption{Inference on the central position of the gaussian distribution of speakers that have been turned on, \mul. The true values of all relevant parameters is provided in the panel titles. The top (bottom) panel illustrates a case where $\mul<\xmax$ ($\mul>\xmax$). The  \if\blackbgd1 cyan dot-dashed \else blue dot-dashed \fi and  \if\blackbgd1 orange dashed  \else cyan dashed  \fi  curves are the posteriors obtained either by numerical or analytical average of the likelihood samples. The yellow curves represent likelihood of individual measurements whereas the thick black (thin blue) vertical line represents $x_\mathrm{max}$ ($\mut$). Finally, the black solid curve (right y-axis) is the fraction of detectable sources, Fig.~\ref{Fig.Alpha}. The cut of the x-axis at $-2$m is just for graphical reasons, and not a cut on the sources or samples used. } \label{Fig.Speakers1}
\end{figure}

The top panel of  Fig.~\ref{Fig.Speakers1}  shows an example of the inference on \mul obtained with 150 sources. In this case the true mean of the population prior, \mut, is smaller than the threshold \xmax.
The 150 individual likelihoods are shown in \if\blackbgd1 ghostly green \else yellow \fi$\!\!$. The thick vertical line marks \xmax.
It is worth stressing again how it is entirely possible and perfectly fine that samples from individual events extend to values larger than \xmax: \emph{all} those samples must be used to calculate the numerator of Eq.~\refp{Eq.MulPostMean}.
The \if\blackbgd1 green \else black \fi line (right y-axis) is the detection efficiency $\alpha(\mul)$, already shown in \Cref{Fig.Alpha}.
Finally, the  \if\blackbgd1 cyan dot-dashed \else blue dot-dashed \fi and  \if\blackbgd1 orange dashed  \else cyan dashed  \fi curves are the posteriors on \mul obtained either with the fully numerical method, Eq.~\refp{Eq.MulPostMean}, or the analytical marginalization over the samples, Eq.~\refp{Eq.MulFinal}, respectively

It is worth stressing that this inference can be carried on even if most of the speakers are not detectable, bottom panel of \Cref{Fig.Speakers1}. In this case $\mut=8.3$ and only $\sim 7\%$ of the speakers produce a detectable sound.
Studying the figure it might at first look surprising that the posterior peaks at the true value, in a region where no data is available, if not the tail of the distribution of the detectable events.
What is happening is that, as seen above, each time we try a value of $\mul$ we need too divide by $\alpha(\mul)$ \emph{to the power \Ntri}. Thus, since $\alpha(\mut)\simeq 0.07$ while $\alpha(\xmax)=0.5$, for example, the product of the tails of the likelihoods at \mut is enhanced by a factor of $(0.5/0.07)^{150} = 1.2\times 10^{128}$ (!) relative to the product of the likelihoods at \xmax.

The fact that inference can be carried out correctly even if the population is such that most events cannot be detected should not be surprising. In fact, it is quite common. In GW astrophysics, the signal-to-noise ratio ---which is the detection criterium\footnote{At least to first order; the real case is more complicated; see below.}--- decreases as $1/d_L$, where $d_L$ is the source luminosity distance. If sources are placed with a distance distribution that goes like $d_L^\beta$, then the true value of $\beta$ is such that most events are in fact undetectable.\footnote{It is worth checking out the open-source code by W. Farr which was prepared as supplementary material of ~\cite{Mandel:2018mve}:  \url{https://github.com/farr/SelectionExample}.}

\section{A gravitational-wave example} \label{Sec.GWExample}

In this section we are going to look at a more complicated example from the research field of the authors: population inference on the mass distribution of binary black holes. We will assume that  all detected binary black holes emitting GWs are members of the same population, and that their masses are draws from Gaussian distributions of unknown means. While the details of the model are discussed below, let us stress here that we are considering a single population. In general, it might be the case that more than one population exists (e.g. some of the black hole binaries might have formed in galactic fields and others in globular clusters). This can be dealt using a model for each of the populations, and treat the number of sources produced in each as one of the unknowns being measured from the data.\footnote{More often, one works with \emph{branching ratios}, i.e. the ratio of sources produced in any pair of populations.} If enough sources are detected, and if the various populations have some distinguishing features, one will be able to measure both number of sources produced by each population, and their characteristic parameters~\cite{Farr:2015,Vitale:2015tea,Rodriguez:2016vmx,Talbot:2017yur,Farr:2017uvj,LIGOScientific:2018jsj, 2019ApJ...886...25B}. A multi-population analysis would then proceed along the same lines of the one described below, but it would be computationally much more expensive due to the larger number of parameters.
\vskip 0.5cm
We will work under the simplifying assumptions that one GW detector has collected data for a total time $T$, and a search algorithm has been searching for binary black hole coalescences, finding a certain number of triggers. No background exists, and we are not interested on the total number of events, which is marginalized using a Jeffreys prior.  Under these assumptions, the posterior for the population parameters is given by Eq.~\refp{Eq.PostLambdaOfPost}.

In our simple model, black holes have no spins, and they inspiral in quasi-circular orbits. If this is the case, each GW event depends on 9 unknown parameters: the two masses ($m_1$ and $m_2$)
, the right ascension and declination (RA, dec), the orbital orientation ($\iota$) and polarization ($\psi$), the arrival time and phase ($t_c$, $\varphi_c$) and the luminosity distance\footnote{In practice, we will use the redshift $z$ instead, converting the one to the other using the Plank 2015 cosmology~\cite{2015arXiv150201589P} as coded up in \texttt{astropy}~\cite{astropy:2018}.} $d_L$ \cite{2015PhRvD..91d2003V}. Collectively, these 9 parameters indicated as \vt.

We will assume that noise is Gaussian, stationary and additive, such that in presence of an astrophysical signal $h(\vt)$ the recorded time-domain data is:
\beq
d(t)= n(t) + h(t,\vt)\,,
\eeq
the Gaussianity of the noise naturally leads to a Gaussian likelihood for the data:
\beq
p(d | \vt) \propto e^{- <d - h(\vt) | d - h(\vt)>}\,,
\eeq
where we have defined the inner product in the frequency domain:
\beq
\langle a(f) | b(f) \rangle  \;\equiv 2 \int \ud f \frac{a(f)\, b^*(f) + a^*(f)\,b(f)}{S(f)}\,,
\eeq
and $S(f)$ is the noise auto-correlation, aka noise power spectral density (PSD)~\cite{1987thyg.book..330T,Martynov:2016fzi,2016LRR....19....1A,SathyaSchutzLRR}. This function is small (large) where the detector is more (less) sensitive.
The integration range should be wide enough to cover the frequency content of the eventual signal, and typically is limited on the low side by the seismic wall of ground-based detectors ($\sim 10-20$~Hz), and on the high side by the Nyquist frequency.

We need to calculate the fraction of detectable events $\alpha(\vl)$ given some hyper population:
\beq
\alpha(\vl)= \int \ud\vt\; p(\dete | \vt) \pi(\vt |\vl)\,.
\eeq
Given significant computational resources and patience, one would evaluate this integral by first calculating $p(\dete | \vt)$  using the actual output of a search algorithm on a large set of simulated signals, and then doing appropriate reweighting according to the population prior~\cite{LIGOScientific:2018jsj}.

For our simple example, and for a significant fraction of the literature, a simpler approach is enough, which strongly reduces the amount of required computing time.
First, we need to decide on a detection criterium which can be easily verified without having to run a full search for compact binaries. We introduce the optimal signal-to-noise ratio (SNR) $\rho$ of a GW signal with parameters \vt  as
\beq
\rho^2 \equiv \langle h(f,\vt) | h(f,\vt)\rangle\,,
\label{Eq.OptSNRDef}
\eeq
where $h(f,\vt)$ is the Fourier transform of the GW signal, and the inner product is defined above.
We can call a signal detectable if it gives an optimal SNR larger than a threshold \rhot (usually 8 or so) in a single detector. As mentioned 	above, reality is more complicated, and in practice a few more factors will contribute to making a signal detected by a real GW search algorithm~\cite{Sachdev:2019vvd,Usman:2015kfa,LIGOScientific:2018jsj,GW150914-CBC,2020arXiv200706585G}, but this is a good first approximation~\cite{Wysocki:2018mpo,LIGOScientific:2018jsj}.

To explicitly calculate $\alpha(\vl)$, we need to say a bit more about the actual GW signal and how it depends on \vt. As mentioned above, we will make the simplifying assumptions that no spins are present (In general this won't affect the SNRs by more than a few percent~\cite{Ng:2018neg,2018PhRvD..98h4036G}). Furthermore, we will work in the limit where only the main multipolar emission mode  described by $\ell=m=2$ ~\cite{Maggiore} dominates, which is reasonable for binaries of stellar-mass black holes with roughly equal masses~\cite{1994PhRvD..49.2658C,LIGOScientific:2020stg}. In this scenario, there is a clean separation in the GW amplitude between extrinsic and intrinsic parameters:
\beq
h(f,\vt)= w(\text{RA,dec},\iota,\psi)\frac{g(m_1,m_2 ,z,f)}{d_L(z)} \sqrt{\frac{5}{24 \pi^{4/3}}} e^{i \psi(f,\vt)} \,,%
\eeq
where the function $w$ is defined as~\cite{1993PhRvD..47.2198F,1996PhRvD..53.2878F,1994PhRvD..49.2658C,Flanagan:1997sx,OShaughnessy:2009szr,Dominik:2014yma} :
\beq
w=\sqrt{\left(\frac{1+\cos ^{2} \iota}{2} {F}_{+}\right)^{2}+\left(\cos \iota {F}_{\times}\right)^{2}}\,,
\eeq
where
$F_+(\text{RA,dec},\psi)$ and $F_\times(\text{RA,dec},\psi)$ are the antenna patters for the two GW polarizations: they are large in regions of the sky where the GW detectors are more sensitive~\cite{SathyaSchutzLRR,2011CQGra..28l5023S,1989JBAA...99..196H}.
The function \w is  easy to calculate, and takes values in the range $[0,1]$.
The maximum value of \w is reached for a CBC source that is exactly above the detector and with inclination equal to zero (i.e. angular momentum parallel to the line of sight).

If one refers to this lucky source as \emph{optimally oriented}, then the SNR of any other source with the same masses and distance, but different sky position and orientation, can be re-cast in terms of the optimally oriented SNR:

\beq
\rho(\vt)= w(\text{RA,dec},\iota,\psi)\, \rho_{oo}(m_1,m_2,z)
\eeq
where ``oo'' stands for optimally oriented.

We can write the detection efficiency as:
\begin{align}
\alpha(\vl)&= \int \ud m_1 \ud m_2 \ud z \bigg[\int  \ud(\text{RA,dec},\iota,\psi) \notag\\ &\times p(\dete | \text{RA,dec},\iota,\psi,m_1,m_2,z) \pi(\text{RA,dec},\iota,\psi)\bigg] \pi(m_1 m_2 z |\vl) \nonumber.
\end{align}
where we have made the assumption that the population priors for the sky position and orientation are independent from the ones for masses and redshift. We am using $ \ud(\text{RA,dec},\iota,\psi)$ as a short way of indicating the product of differentials of all 4 variables.

Let us focus on the integral in square brackets. It contains the probability that an event is detected, given the value of all parameters~\footnote{You might have noticed that the coalescence time and phase do not appear in this expression. This is because they do not contribute to the optimal SNR, hence integrating over them will just given an overall factor to $\alpha$, the same for all events and all \vl, which can thus be neglected.}.
Given our criterium for detectability, this is exactly 1 if \vt is such that the SNR is above \rhot, and zero otherwise.
The integral can thus be written as
\begin{align}
\int  \ud(\text{RA,dec},&\iota,\psi) \mathrm{H}(\rho- \rhot) \pi(\text{RA,dec},\iota,\psi) \notag \\ &= \int  \ud(\text{RA,dec},\iota,\psi) \mathrm{H}(\w \rho_{oo}- \rhot) \pi(\text{RA,dec},\iota,\psi)  \notag \\ &=
\int  \ud(\text{RA,dec},\iota,\psi) \mathrm{H}\left(\w - \frac{\rhot}{\rho_{oo}}\right) \pi(\text{RA,dec},\iota,\psi)\,,\label{Eq.AngleInt}
\end{align}
with H the Heaviside function. Note that all the dependence on the masses and distance is now contained in $\rho_{oo}$.
Solving this integral can be done numerically as follows:

\begin{itemize}
\item Select a population prior for the sky position and orientation. The standard choice is isotropic on both. That gives $ \pi(\text{RA,dec},\iota,\psi) = 1/4 \pi^4$
\item Generate a large number of samples from this prior, and calculate \w for all of them. Those points represent a probability density function $p(\w)$.
\item The integral above becomes:

\beq
\int_{\frac{\rhot}{\rho_{oo}}}^1 \ud \w\; p(w)= 1 - \mathrm{CDF}_\w\left(\frac{\rhot}{\rho_{oo}}\right)\,,
\label{Eq.CCDFIntDef}
\eeq
where CDF is the usual cumulative distribution function of a variable.

This can also be seen as follows: let us pack $\text{RA,dec},\iota,\psi$ into a vector $\vec{y}$ then Eq.~\refp{Eq.AngleInt} becomes:
\beq
\int  \ud\vec{y}\; \mathrm{H} \left(\w (\vec{y}) - k\right) \pi(\vec{y})= \frac{1}{\Nsa} \sum_{j=1}^{\Nsa} \left. \mathrm{H} \left(\w(\vec{y}\,^j) - k\right) \right |_{\vec{y}^j \sim \pi(\vec{y})}\,,
\eeq
where we called $k\equiv {\rhot}/{\rho_{oo}}$ which is a constant when this integral is evaluated. The term inside the sum is either 1 or 0, depending on whether the $\vec{y}$ value being draw from the prior results in $\w(\vec{y})$ to be larger or smaller than $k$. That is, Eq.~\refp{Eq.AngleInt} is the fraction of points drawn from  $\pi(\vec{y})$ [i.e. $\pi(\text{RA,dec},\iota,\psi)$]  which gives a \w larger than some number $k$. This is exactly what Eq.~\refp{Eq.CCDFIntDef} shows.

One minus the CDF, is often called the complementary cumulative distribution function (CCDF) or tail function. We will use CCDF from now on:
\beq
\int_{\frac{\rhot}{\rho_{oo}}}^1 \ud \w\;  p(w)\equiv \mathrm{CCDF}_\w\left(\frac{\rhot}{\rho_{oo}}\right)\,.
\eeq

\begin{figure}[t]
\centering
\if\blackbgd1
\includegraphics[width=0.8\textwidth]{wfactor_bw.pdf}
\else
\includegraphics[width=0.8\textwidth]{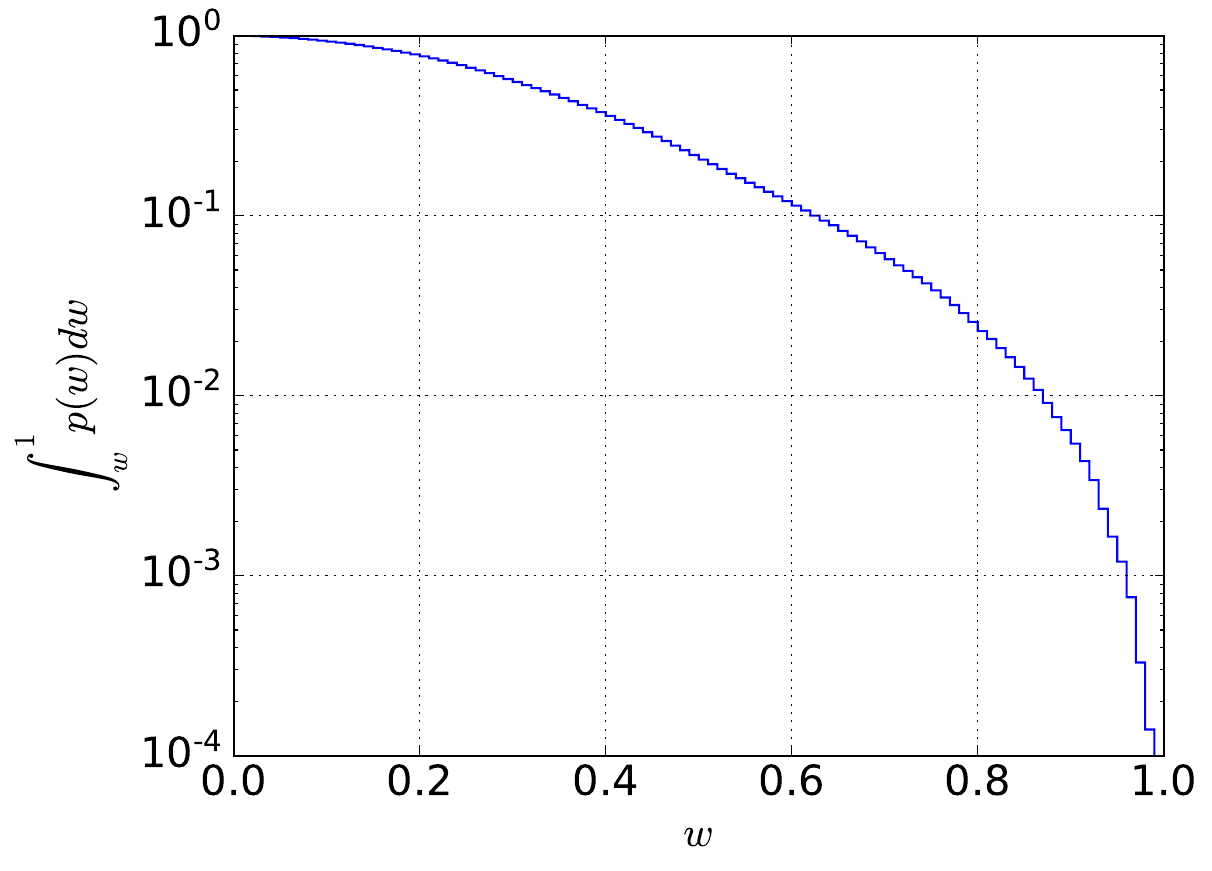}
\fi
\caption{The complementary cumulative function for $w$.}
\end{figure}

We have reduced all of the dependence on masses and distance to restricting the range of a 1D integral! We typically just calculate the distribution of \w once, and then calculate the CCDF, or rather use an interpolator which makes it even less time consuming, as many times as we want during the analysis.

Back to the selection function, we can now write it as

\boxit{
\alpha(\vl)= \int \ud m_1 \ud m_2 \ud z\;  \mathrm{CCDF}_\w\left(\frac{\rhot}{\rho_{oo}}\right) \pi(m_1 m_2 z |\vl).
}{AlphaSimple}
\end{itemize}

\begin{figure}[t]
\centering
\if\blackbgd1
\includegraphics[width=0.8\textwidth]{PdetAtZPointOne_bw.pdf}
\else
\includegraphics[width=0.8\textwidth]{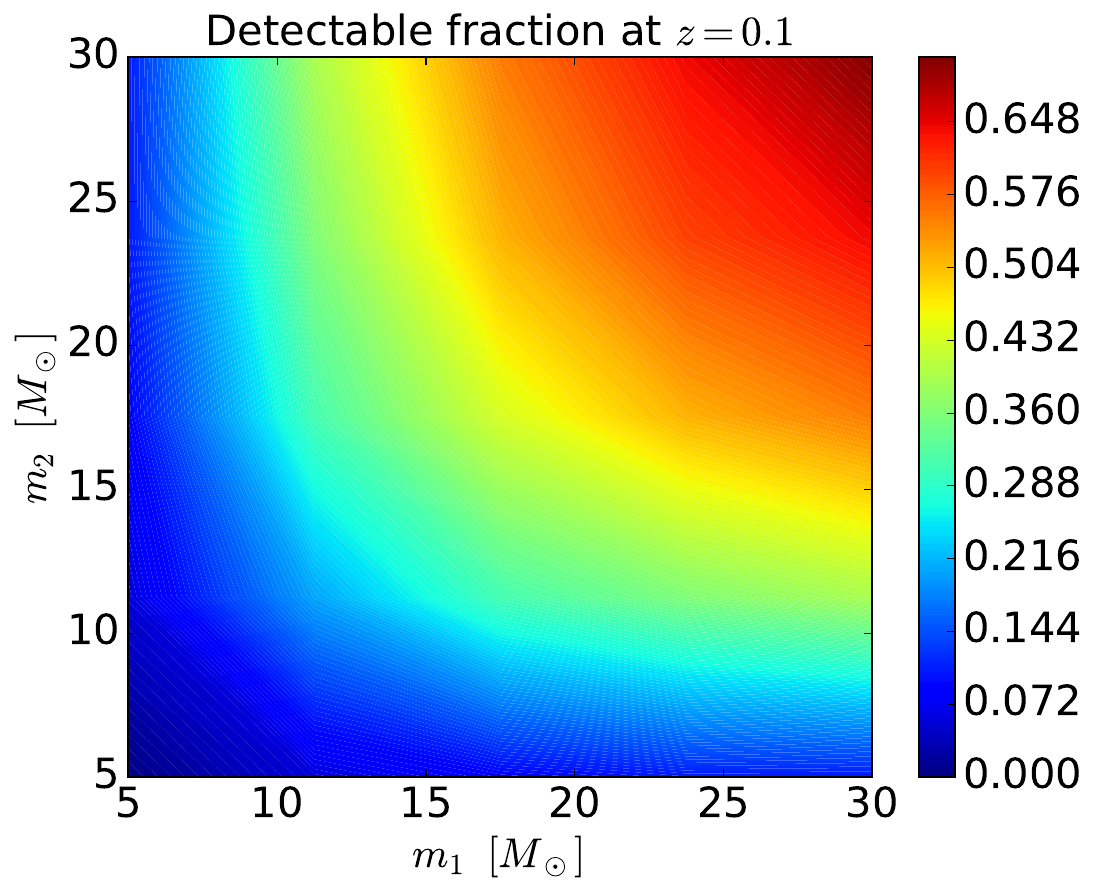}
\fi
\caption{The fraction of detectable BBH of given $m_1$ and $m_2$ (in the source frame) at $z=0.1$. A single interferometer is used with a sensitivity representative of LIGO's second observing run~\cite{2016LRR....19....1A}, together with a threshold SNR of 8. Notice that in most of the GW literature one assumes by default that $m_1\geq m_2$.}\label{Fig.PdetOfMasses}
\end{figure}

The complementary cumulative distribution function is a function of the two masses and the redshift, and it can be interpreted as the fraction of sources with given masses and redshift that are detectable.  \Cref{Fig.PdetOfMasses} shows this probability as a function of the two (source-frame) masses, fixing the redshift $z=0.1$.

We have reduced the problem, within our simplifying hypothesis, to a 3D integral where the only computationally consuming part is the calculation of the SNR of an optimally oriented source of given masses and redshift.
In practice, this can also be made very fast using interpolation in the mass plane, and the fact that the SNR scales linearly with the distance.
There is a catch though: the SNR depends on the detector-frame masses, which are redshifted by a factor of $1+z$ over the astrophysical masses, the ones we care about ~\cite{1986Natur.323..310S,Krolak1987,Maggiore}. The integral for $\alpha$, as written above, is on the source-frame masses.
The simple rescaling of the SNR as $1/\dl$ would give us the SNR of a source with the same \emph{detector-frame} masses at different distances, if we know its SNR at some reference distance. But all those sources would thus have different \emph{source-frame} masses! This is not a problem, we just need to fold in the redshift carefully when we rescale the SNR of sources.\newline\\
\noindent
This completes our discussion on how to calculate the selection function. If all we want is to infer \vl, and not the number of sources that occurred over the experiment time (or their rate per unit time) we can just use Eq.~\refp{Eq.HyperPost}. Otherwise, some more work is required.
This is because $\alpha$ does not tell us anything about the actual number of events, only about the \emph{fraction} of detectable events.
If we want to do inference on the actual number of sources, or their time-rate, we also need to calculate the expected  number of detectable events $\Nabos$ over the experiment time $T$ so that we can calculate Eq.~\refp{Eq.HyperPosteriorFinal}. Luckily, this is straightforward because \N and \Nabos are related by $\alpha$, which we already have:
\beqa
\Nabos= \N \;\alpha(\vl) &=& \int \ud m_1 \ud m_2 \ud z\;  \mathrm{CCDF}_\w\left(\frac{\rhot}{\rho_{oo}}\right) \N \pi(m_1 m_2 z |\vl)\nonumber\\
&=&  \int \ud m_1 \ud m_2 \ud z\;  \mathrm{CCDF}_\w\left(\frac{\rhot}{\rho_{oo}}\right) \frac{\ud^3 \N}{\ud m_1 \ud m_2 \ud z}\left(\La\right)\,.
\eeqa
where we have used Eq.~\refp{Eq.pTheta} to relate the population prior to the differential rate per unit masses and unit redshift. Usually, the generation of compact binaries is quoted in units of mergers per unit of comoving volume per unit of (source-frame) time, and in the most generic case depends on masses, redshift, spins, as well as environmental factors (e.g. the metallicity). Let us use the symbol \vr for it and only explicit report masses and redshift dependence:
\beq
\vr(m_1,m_2,z) \equiv \frac{\ud^2 \N}{\ud V_c \ud t_s}\,.
\label{Eq.PopSynRate}
\eeq
One has
\beqa
\Nabos&=&  \int \ud m_1 \ud m_2 \ud z\;  \mathrm{CCDF}_\w\left(\frac{\rhot}{\rho_{oo}}\right) \frac{\ud^4 \N}{\ud m_1 \ud m_2 \ud V_c\ud t_s} \frac{\ud V_c}{\ud z} \frac{\ud t_s}{\ud t_d}\ud t_d\nonumber\\
&= &T \int \ud m_1 \ud m_2 \ud z\;  \mathrm{CCDF}_\w\left(\frac{\rhot}{\rho_{oo}}\right) \frac{\ud^2 \vr}{\ud m_1 \ud m_2}\left(m_1,m_2,z,\La\right) \frac{\ud V_c}{\ud z} \frac{1}{1+z} \label{Eq.VTExpl}\,.
\eeqa
where we have used the standard cosmological time dilation formula ${\ud t_s}/{\ud t_d}=1/(1+z)$ to relate time measured in the detector frame, $t_d$, and time measured in the source frame, at redshift $z$, $t_s$~\cite{2003gieg.book.....H}. By definition, the integration over the detector-frame time gives the total observing time $T$ as measured with our clocks.

Equation~\refp{Eq.VTExpl} is fundamentally the product of three terms: an observing time, a source rate per unit time per unit volume, and a volume. What makes it look more complex than this is the fact that the source rate in general depends of a few parameters, and the fact that only part of that rate will contribute to the detectable signals.

Before we continue with our example, we would like to report two alternative ways of writing this same equation, which are often found in the literature. One is (e.g. ~\cite{Dominik:2014yma,2018arXiv180510270F,LIGOScientific:2018jsj}):

\beqa
\Nabos&=& T \int \ud m_1 \ud m_2 \ud z\; p_\mathrm{det}(m_1,m_2,z) \frac{\ud^2 \vr}{\ud m_1 \ud m_2}\left(m_1,m_2,z,\La\right) \frac{\ud V_c}{\ud z} \frac{1}{1+z}\,,
\eeqa
where $p_\mathrm{det}(m_1,m_2,z)$ is used for the probability, averaged over sky positions and orientations, that a source of given source frame masses and redshift is detectable.
The other makes use of the so called the \emph{sensitive} volume~\cite{GW150914-ASTRO,Chen:2017wpg}:
\beq
V(m_1,m_2) \equiv \int \ud z \; \frac{1}{1+z} \frac{\ud V_c}{\ud z}  p_\mathrm{det}(m_1,m_2,z)
\eeq
to define the \emph{sensitive spacetime volume}%
\beq
VT(m_1,m_2)
\eeq
with $T$ being the observation time measured in the detector frame. The sensitive time-volume, usually reported in $\mathrm{Gpc^3\; yr}$, represents the amount of universe spacetime volume ``covered'' by the instrument in searching merging black-holes of masses $(m_1,m_2)$, after an observing time $T$, as measured by the instrument clocks. Then, if the intrinsic merger rate \vr is assumed constant in time and redshift,  the number of expected detectable sources after an observing time T can be written~\cite{Wysocki:2018mpo,GW150914-ASTRO,LIGOScientific:2018jsj}:

\beq
\Nabos = \int \ud m_1 \ud m_2   \frac{\ud^2 \vr}{\ud m_1 \ud m_2}\!\left(m_1,m_2,\La\right) \,VT(m_1,m_2)
\eeq

\vskip 0.5cm

We now have all the background information to approach our problem. We will assume that the source-frame masses of the BHs in the binaries come from independent gaussian distributions of \emph{known} standard deviations (equal to 5 solar masses (\msun)~ for both $m_1$ and $m_2$) and unknown means $\mu_1$ and $\mu_2$, which we want to infer. Our hyper parameter vector is thus $\vl=\{\mu_1,\mu_2\}$, and we will use a uniform hyper prior $\pi(\mu_1)=\pi(\mu_2) = (35 - 5)^{-1} \msun^{-1} $. In more realistic examples, little is known and the problem's dimensionality can be very high (for a state-of-the-art analysis see e.g. Ref.~\cite{LIGOScientific:2018jsj}).
Sky position and orientation of the sources are isotropic, and we assume that sources are distributed uniformly in comoving volume.

The only parameters that depend on a hyper level are thus the two component masses. We take their population prior to be:
\beq
\pi(m_1 m_2 | \vl\equiv \{\mu_1,\mu_2\})= \mathcal{N}(m_1- \mu_1,5~\msun)  \; \mathcal{N}(m_2- \mu_2,5~\msun)\,,\label{Eq.PopPriorMs}
\eeq
where as before $\mathcal{N}$ is the normal distribution. Thus the population prior we will use when calculating Eq.~\refp{Eq.AlphaSimple} reads:
\beq
\pi(m_1 m_2  z | \vl) = \pi(m_1 m_2 | \vl) \;\pi(z) =  \frac{\ud V_c}{\ud z}(z) \;\mathcal{N}(m_1- \mu_1,5~\msun)  \; \mathcal{N}(m_2- \mu_2,5~\msun)\,,
\eeq
where $\pi(z)={\ud V_c}/{\ud z}$ is the prior on the redshift that results from a uniform in comoving volume prior.

Next, we need to generate some \emph{detectable} sources from the true population, that will be the input used in our inference. If we were applying our analysis to real GW detections, this step would be unnecessary since nature would have produced them for us!
But for this example	 we have to create our own set of detections and the resulting measurements for those parameters.

To generate a set of noisy detections we proceed as follows:\footnote{The procedure described in this page is a simplified version of Appendix A of ~\cite{Fishbach:2019ckx}. It is simpler because we work with uniform priors so that samples from the likelihood and samples from the posterior are equivalent. See also~\cite{2017MNRAS.465.3254M}.}

\begin{itemize}
\item We draw a random value of \vt from the the true population distribution of masses, distance, etc. We take the \emph{true} value of the hyper-parameters to be $\mu^T_1=19~\msun$ and $\mu^T_2=20~\msun$.
\item We calculate the true SNR $\rho_{\rm true}$ of the source.
\item Now, if we want to do a realistic analysis, we cannot quite use \emph{this} SNR to decide if the source is detected, because we measure a noisy version of the true signal. The inferred value of the parameters need not be centered at the true values, but might be offset by an amount that depends on how loud the signal is. We thus add to the true SNR a random number from $\mathcal{N}(0,1)$.\footnote{Given how we have defined SNR above, one can show that if the GW noise were perfectly Gaussian and stationary, the measured SNR would indeed be distributed like this \cite{SathyaSchutzLRR}.}. %
 This gives an observed SNR:
\beq
\rho_{\rm obs}= \rho_{\rm true}+ \mathcal{N}(0,1)\,.
\eeq
\item If $\rho_{\rm obs}> 8$ the source is detected and its \vt recorded.
\item We restart from point 1 and keep going till we generate 20 detectable systems.
\end{itemize}

 Finally, we need to generate for each event a set of samples from the likelihood, which we will need in  Equation~\ref{Eq.PostLambdaOfPost}. This is done as follows:

\begin{itemize}

\item The measured maximum likelihood value of the generic parameter $x$ is taken to be the true value generated above plus an offset $\mathcal{N}(0,\sigma_x)$, where $\sigma_x$ depends on the observed SNR (with louder events having smaller offsets). In a scientific analysis, one would use a simulation campaign to ``tune'' the value of $\sigma_x$ to something representative for each parameters~\cite{2017MNRAS.465.3254M}. In our case we will just use
\beq
\sigma_{m_1}= \frac{1 \msun}{\rho_{\rm obs}}\;;\quad \sigma_{m_2}= \frac{2 \msun}{\rho_{\rm obs}}\,.
\eeq
\item For each event and each parameter we create a set of 150 samples from a gaussian likelihood centered at the \emph{measured} maximum likelihood point, and with a standard deviation equal to $\sigma_x$.\footnote{The variance $\sigma_x$ thus plays a double role: it sets the offset of the measured maximum likelihood value compared to the true value and it sets the width of the likelihood distribution.}
\end{itemize}

It should be stressed that this implies we are assuming the likelihood of each event can be written as the product of a univariate Gaussian distribution for each parameter. This is an obvious over-simplication, which we can afford because we are only after a pedagogical example. In we were working with real samples produced by analyzing real data, the likelihood would in general be more complicated and show non-trivial correlations, but the inference would proceed along the very same steps we are following.

If the population hyper parameters only affect the mass distribution, and as long as the population prior on the other parameters (e.g, redshift) matches what we use to create the population of detected signals (remember Sec.~\ref{SuSec.WhatWeDontSay}) we actually only need to generate likelihood samples for the mass
parameters.

\subsection{What happens next?}

\begin{figure}[b]
\centering
\if\blackbgd1
\includegraphics[width=0.8\textwidth]{Gaussian_post_bw.pdf}
\else
\includegraphics[width=0.8\textwidth]{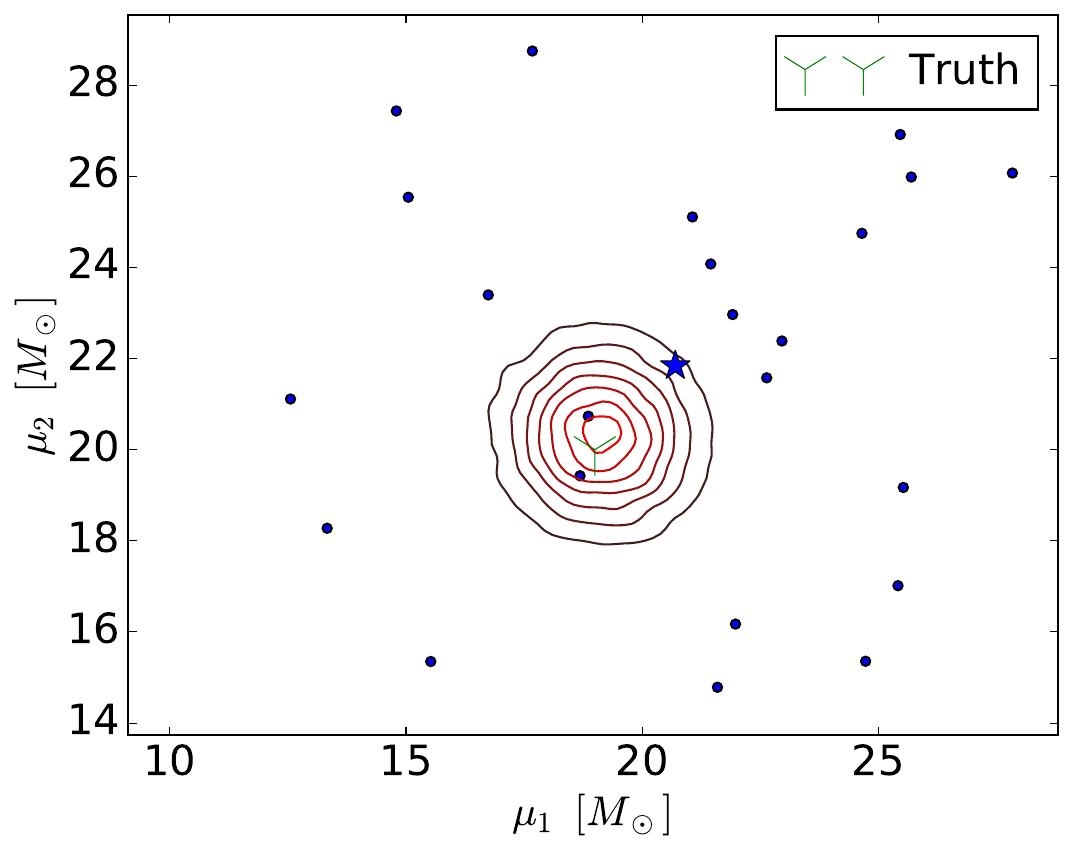}
\fi
\caption{The posterior on the hyper parameters $\mu_1$ and $mu_2$ (red contours) using 20 BBHs. See the body for more details.}\label{Fig.GaussianPost}
\end{figure}

We now have all we need! \Cref{Fig.GaussianPost} shows the main results. The measured maximum likelihood mass values of the individual 20 detected sources are shown as small circles. The \if\blackbgd1 green \else blue \fi star shows what one would obtain if they just took the mean of these points, without properly taking into account selection effects, which makes heavier systems easier to measure (cf. Fig.~\ref{Fig.PdetOfMasses}). As one could have guessed, the naive mean is at larger values of mass than the truth.
Finally, the contours show the proper posterior on the hyper parameters obtained using Eq.~\refp{Eq.PostLambdaOfPost}, which is centered at the true value $(\mu^T_1,\mu^T_2)$.

\begin{figure}[htb]
\centering
\if\blackbgd1
\includegraphics[width=0.95\textwidth]{Gaussian1DDraws_bw.pdf}
\else
\includegraphics[width=0.95\textwidth]{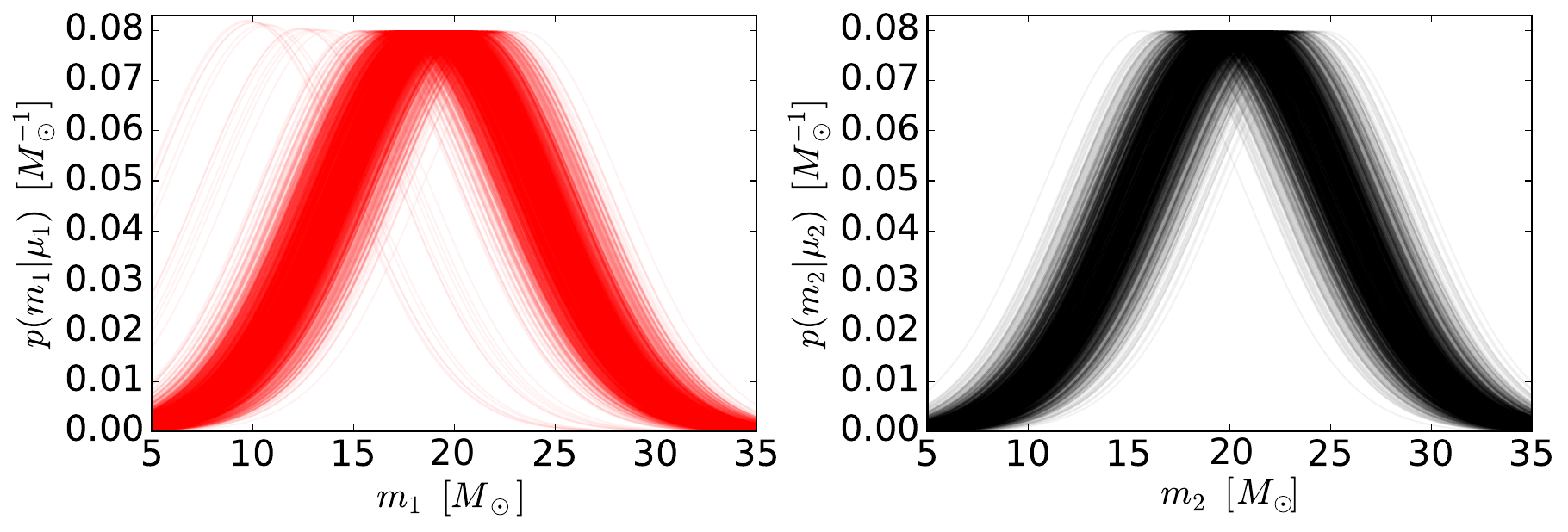}
\fi
\caption{(left:) Probability $p(m_1 | \mu_1)$ for 2000 draws from the posterior distributions of $\mu_1$; (right:) same, but for the other mass. %
}\label{Fig.Gaussian1DDraws}
\end{figure}
\begin{figure}[htb]
\centering
\if\blackbgd1
\includegraphics[width=0.95\textwidth]{Gaussian1DPost_bw.pdf}
\else
\includegraphics[width=0.95\textwidth]{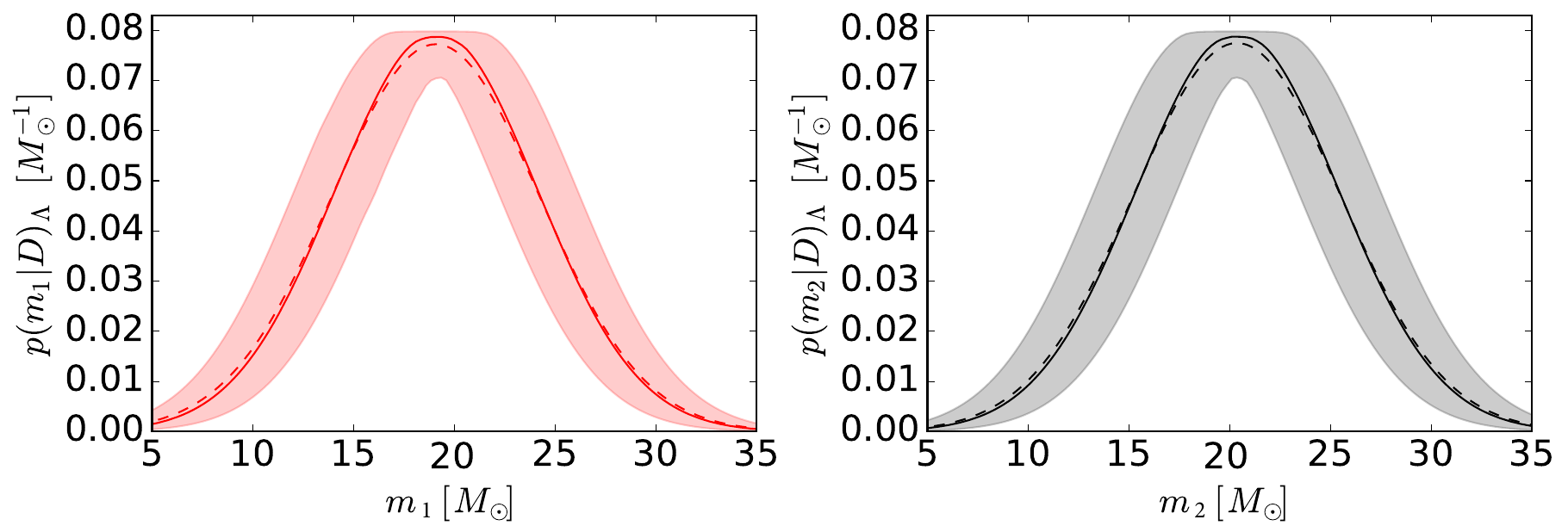}
\fi
\caption{(left:) Median (solid line) and 90\% symmetric credible interval (shaded area) of the 2000 curves in Fig.~\ref{Fig.Gaussian1DDraws} for $m_1$. The dashed line is the posterior predictive distribution; (right:) same, but for the other mass.
 }\label{Fig.Gaussian1DAll}
\end{figure}

What did we learn? There are at least a few different way of answering this question.  The most straightforward answer is: ``we now have a measure of the hyper-parameters, including their uncertainty". This is what was shown by the contours of \Cref{Fig.GaussianPost}.
But we can actually do more. With the samples from the hyper posterior in hand, we can visually represent draws from the population prior, Eq.~\refp{Eq.PopPriorMs}. This is shown in Fig.~ \ref{Fig.Gaussian1DDraws}. We draw 2000 samples from the 1D marginalized posterior for $\mu_1$, $p(\mu_1 | D)$, for each of them we calculate $p(m_1 | \mu_1)$ and plot it in the left panel. The right panel shows the same set of lines for $m_2$.

This figure is objectively quite nice and hypnotic %
but having so many lines makes it hard to interpret. What we can do instead, is to e.g. calculate the 5th and 95th percentile, together with the median, of all draws at every point in the x axis, and plot these instead.
This is show for both masses in Fig.~\ref{Fig.Gaussian1DAll}. The solid curve represents the median of all the lines in Fig.~\ref{Fig.Gaussian1DDraws} at each point, and the shaded areas show the symmetric 90\% credible interval.
These curves represent what we know about the distribution of the component masses in light of our 20 triggers.

A related quantity one often uses is the posterior predictive distribution (PPD)~\cite{gelmanbda04}.
The PPD of a generic parameter vector \vt is the posterior we assign to the parameters in light of the data we observed:
\beq
p( \vt | D)_\La\,,
\eeq
where we have put a subscript \La to stress that this is \emph{not} the posterior of  \vt given the data (which would not necessarily require any knowledge about the population). The PPD tells us what we expect the (true) distribution of the parameters to be, in light of the observation we made. As such, it can be used to quantitatively predict (hence the name) what \emph{future} observations
for sources from our population will look like, given the ones we already made and the selection effects of the detector\footnote{Please notice that $p( \vt | D)_\La$ is the distribution of \emph{true} parameters of the sources \emph{before} selection effects. The distribution of \emph{detected} sources will be different due to selection effects and noise.}

Mathematically, we can related the PPD to the hyper posterior samples we have acquired by using the sum rule:
\beq
p( \vt | D)_\La = \int  \ud\La \; p( \vt | D  \La) p(\La | D)= \int  \ud\La \; \pi( \vt |  \La) p(\La | D)\,,
\eeq
where we have used the fact that the distribution of \vt is fully specified if \La is known to remove the data from the right hand side of the condition bar in $ p( \vt | D  \La)$, which thus is simply the population prior $\pi( \vt |  \La)$.

Given the \Nsa we already got from the hyper posterior, $p(\La | D)$, we can approximate this as:
\beq
p( \vt | D)_\La \simeq \frac{1}{\Nsa} \sum_{i=1}^{\Nsa} \left. \pi( \vt |  \La^i) \right|_{\La^i \sim p(\La | D)}\,.
\eeq
In other words, we calculate the population prior at each value of the hyper posterior samples and take the mean. The marginalized 1D PPD for $m_1$ and $m_2$ in our problem are thus just  the mean of the curves shown in \Cref{Fig.Gaussian1DAll}, and we report them in \Cref{Fig.Gaussian1DAll} using dashed lines.

\section{Conclusions}

As the number of detected sources increases (in GW astronomy much like in other experimental sciences), the set of source can be used to shed light on the underlying astrophysical populations. In this Chapter, we provided a pedagogical derivation of the main equations one needs to perform Bayesian hyperparameter inference. The key result is given in Eq. (\ref{Eq.HyperPost}), which is the population posterior used in current state-of-the-art GW applications. We hope that our numerous remarks on the role of hypothesis and the meaning of the notation have clarified some of the key concepts in this increasingly relevant branch of statistical physics.

\section*{Acknowledgments}

We thank Sylvia Biscoveanu for a critical reading of an early draft of these notes, as wells as Emanuele Berti, Maya Fishbach, Max Isi and Ken Ng for long and illuminating discussions. We thank Nancy Aggarwal and Peter Couvares for useful comments. 
S.V. acknowledges the support of the National Science Foundation and the LIGO Laboratory and the support of the MIT physics department through the Solomon Buchsbaum Research Fund.
D.G. is supported by European Union's H2020  ERC Starting Grant No. 945155--GWmining, Leverhulme Trust Grant No. RPG-2019-350, and Royal Society Grant No. RGS-R2-202004.
LIGO was constructed by the California Institute of Technology and Massachusetts Institute of Technology with funding from the National Science Foundation and operates under cooperative agreement PHY-0757058.
We acknowledge use of \texttt{iPython}~\cite{ipython}, \texttt{Matplotlib}~\cite{Hunter:2007,matplotlib}, \texttt{NumPy}~\cite{numpy2}, \texttt{SciPy}~\cite{scipy}, \texttt{emcee}~\cite{2013PASP..125..306F} and \texttt{SeaBorn}~\cite{michael_waskom_2017_883859}.
This is LIGO Document P2000231

\begingroup\let\newpage\relax
\glsaddall
\printglossary[type=main,nonumberlist,title={\glname}]
\endgroup

\section*{Cross-References}

{\color{red} Include a list of related entries from the handbook here that may be of further interest to the readers.}

\bibliographystyle{apsrev}
\bibliography{Notes}

\end{document}